\documentclass[onecolumn,10pt,superscriptaddress,notitlepage,nofootinbib,floatfix,prx,aps]{revtex4-2}
\usepackage[usenames,dvipsnames]{xcolor}
\usepackage{mathtools}
\usepackage{multirow}
\usepackage[colorlinks, linkcolor=blue,anchorcolor=blue,citecolor=blue,urlcolor=blue]{hyperref}
\usepackage{amssymb}
\usepackage{pifont}
\newcommand{\cmark}{\ding{51}}

\usepackage{physics}
\usepackage{natbib}
\setcitestyle{super}
\usepackage{xspace}
\usepackage{array}
\usepackage{tabularx}
\newcolumntype{M}{>{\centering\arraybackslash}X}
\usepackage{array}
\usepackage{tabularx}
\usepackage{float}
\usepackage{booktabs}
\usepackage{enumitem}
\usepackage{makecell}
\usepackage{colortbl}
\newcolumntype{M}{>{\columncolor{blue!7}}r}
\usepackage[multiple]{footmisc}

\newcommand{\strong}{\textcolor{OliveGreen}{$\checkmark$}}
\newcommand{\partialok}{\textbf{\textcolor{RedOrange}{$\triangle$}}}
\newcommand{\notfocus}{$\square$}

\usepackage[most]{tcolorbox}

\newcommand{\NA}{{--}}

\newcommand{\mgym}{\texttt{metriq-gym}\xspace}
\newcommand{\mdata}{\texttt{metriq-data}\xspace}
\newcommand{\mweb}{\texttt{metriq-web}\xspace}
\newcommand{\metriq}{\texttt{metriq.info}\xspace}
\newcommand{\qbraid}{\texttt{qBraid}\xspace}
\newcommand{\ufrepo}[1]{\href{https://github.com/unitaryfoundation/#1}{\texttt{#1}}}
\newcommand{\hreftexttt}[2]{\href{#1}{\texttt{#2}}}

\begin{document}

\title{Metriq: A Collaborative Platform for Benchmarking Quantum Computers}

    \author{Alessandro Cosentino}
    \thanks{These authors contributed equally to this work. Correspondence should be addressed to metriq@unitary.foundation.}
    \affiliation{Unitary Foundation, San Francisco, CA, USA}
    
    \author{Changhao Li}
    \thanks{These authors contributed equally to this work. Correspondence should be addressed to metriq@unitary.foundation.}
    \affiliation{Unitary Foundation, San Francisco, CA, USA}
    
    \author{Vincent Russo}
    \thanks{These authors contributed equally to this work. Correspondence should be addressed to metriq@unitary.foundation.}
    \affiliation{Unitary Foundation, San Francisco, CA, USA}
    
    \author{Bradley A. Chase}
    \affiliation{Unitary Foundation, San Francisco, CA, USA}
    
    \author{Tom Lubinski}
    \affiliation{Quantum Circuits, Inc., New Haven, CT, USA}
    \affiliation{QED-C Technical Advisory Committee on Standards and Performance Benchmarks}
    
    \author{Siyuan Niu}
    \affiliation{University of Central Florida, Orlando, FL, USA}
    
    \author{Neer Patel}
    \affiliation{University of Central Florida, Orlando, FL, USA}
    
    \author{Nathan Shammah}
    \affiliation{Unitary Foundation, San Francisco, CA, USA}
    \affiliation{Unitary Fund France, Tours, France}
    
    \author{William J. Zeng}
    \affiliation{Unitary Foundation, San Francisco, CA, USA}
    \affiliation{Quantonation, New York, NY, USA}

\date{March 9, 2026}

\begin{abstract}
    The fragmented landscape of quantum computer benchmarks, characterized by
    system-specific tools and inconsistent evaluation methodologies, hinders reliable
    cross-platform performance assessment. We introduce Metriq, an open-source
    collaborative platform for reproducible cross-platform quantum benchmarking 
    that integrates benchmark definition and execution, data collection, and public presentation 
    into a unified workflow. 
    The Metriq benchmark suite spans both system-level metrics that characterize fundamental device
    properties such as entanglement quality, gate performance, and circuit speed, as
    well as application-inspired protocols that assess performance on quantum
    machine learning, optimization, and quantum simulation tasks.
    Benchmarks are chosen to scale with processor size, and the framework incorporates
    cost and resource estimation to support practical evaluation.
    Using Metriq, we collect and publicly release results from more than ten quantum
    computers across multiple hardware vendors, enabling systematic cross-platform
    comparison. The resulting curated dataset also reveals the practical strengths
    and limitations of individual benchmarks, creating a feedback loop that informs
    the ongoing refinement of the suite. 
    To summarize performance across the benchmark suite, we introduce the Metriq Score,
    a composite index aggregating benchmark outcomes.  
    We further present cross-benchmark analyses enabled by the shared dataset and their correlations with hardware calibration metrics. 
    Through open development and data sharing, Metriq provides a practical foundation for reproducible benchmarking of quantum computers as hardware and benchmarking methods continue to evolve.
\end{abstract}

\maketitle

% \clearpage

%------------------------------------------------------------------------------%
\section{Introduction}
\label{sec:introduction}
%------------------------------------------------------------------------------%
Over the past decade, quantum processors have improved rapidly in scale, quality, and accessibility. As the hardware landscape diversifies across multiple platforms such as
superconducting circuits~\cite{arute2019quantum, barends2014superconducting},
trapped ions~\cite{chen2023benchmarking, moses2023race, pino2021demonstration},
and neutral atoms~\cite{bluvstein2024logical, graham2022multi},
systematic and comprehensive quantum computer benchmarking has become both more important and more challenging. 
Effective benchmarking goes beyond isolated metrics and captures the interplay between critical properties including circuit quality, execution speed, and scalability~\cite{proctor2025benchmarking,
lubinski2024optimization, lubinski2023application}.

Despite the growing importance of quantum benchmarking, the landscape remains
fragmented. Various quantum hardware providers employ distinct architectures,
interfaces, and error mitigation strategies, making it challenging to perform
standardized performance evaluations. The dynamic nature of quantum technology
development complicates the establishment of stable reference benchmarks,
resulting in inconsistencies in the way benchmarks are defined, executed, and
compared across different quantum computing platforms. Notably, the absence of
fair third-party comparisons across hardware providers complicates objective
evaluations, highlighting the need for community-driven benchmarking
frameworks~\cite{lorenz2025systematic} that can offer unbiased assessments
compared to results from vendor-run benchmarks. Furthermore, while the community
has proposed many benchmarks in parallel with rapid hardware progress, most
results appear as isolated case studies, with little public benchmark data and
almost no standardized cross-platform datasets. Systematic comparisons of the
same protocol across providers and devices and over time are rare, underscoring
the need for platforms that collect and curate open benchmarking data as a
shared resource.

To bridge this gap, we introduce \emph{Metriq}: an open benchmarking platform
that couples a runner, a dataset, and a web portal. The runner, \mgym, provides
a unified interface for defining and executing benchmarks across heterogeneous
providers and hardware architectures. Benchmark results generated by \mgym are
stored in a versioned, schema-validated dataset (\mdata) and exposed through an
interactive web application (\mweb) that supports filtering, drill-down, and
machine-readable export for further analysis. Together, these components are designed to support continuous, cross-platform benchmarking with transparent provenance.

Metriq differs fundamentally from existing benchmarking efforts. Whereas
vendor-maintained frameworks, such as IBM Qiskit device-benchmarking
tools~\cite{qiskit2025benchmarking} or IQM’s benchmarking
utilities~\cite{iqm2025benchmarks}, are necessarily tied to a single technology
stack, Metriq is developed and maintained by an independent, non-vendor third
party and open-source community, avoiding alignment with any specific hardware
architecture or commercial provider. Government-funded and academic efforts have
also advanced the state of quantum performance assessment: Sandia National
Laboratories’ Quantum Performance Laboratory~(QPL) develops gate-set tomography
and related characterization protocols backed by the open-source pyGSTi
toolkit~\cite{nielsen2020pygsti, proctor2025benchmarking}, while the DARPA
Quantum Benchmarking program, in partnership with the University of Technology
Sydney, targets fault-tolerant resource estimation across application
domains~\cite{darpa2024benchmarking}. These initiatives focus primarily on hardware
characterization or resource estimation for specific architectures, whereas
Metriq emphasizes continuous, cross-platform benchmarking of diverse workloads
across multiple providers. Moreover, in contrast to benchmarking
suites that target a specific layer of the stack or enforce a fixed class of
workloads such as QUARK~\cite{finzgar2022framework} and BACQ~\cite{barbaresco2024bacq}, which focus on
application-driven benchmarks, or SupermarQ~\cite{tomesh2022supermarq}, whose
workload-oriented benchmarks are designed to reflect realistic application
structures, Metriq does not impose a prescribed hierarchy. 

Another key distinction is that Metriq is designed as a continuously updated
benchmarking platform. Rather than producing a one-off snapshot that becomes
obsolete as devices evolve, Metriq supports periodic re-execution of benchmarks
across all supported hardware backends, enabling longitudinal tracking of device
behavior and systematic comparison as providers release new systems or update
calibrations. Equally important, the systematic collection of curated,
cross-platform data exposes the practical strengths and limitations of the
benchmarks themselves, revealing which protocols are most discriminative across
devices, where design assumptions break down on real hardware, and how the suite
should evolve. This feedback loop between data curation and benchmark refinement
is a distinctive advantage of a platform-driven approach. Finally, Metriq is
built to be open and community-driven: all benchmarks, execution tooling, data,
and visualizations are maintained in the open, enabling transparent scrutiny,
extension, and collective governance by the broader quantum-computing community.

Our contributions are fourfold:
\begin{enumerate}
    \item We release \emph{Metriq} as an open-source platform for collaborative
    quantum benchmarking, with public tooling for benchmark execution,
    versioned data curation, and web-based result exploration.
    
    \item We design and implement novel vendor-agnostic benchmarks, including
    new protocols such as the Bell State Effective Qubits (BSEQ) benchmark, and
    extend existing ones to hardware platforms where they have not previously
    been reported.
    
    \item We curate an initial, intentionally opinionated benchmark suite
    spanning system-level and application-inspired tasks, execute it across
    multiple providers, and summarize the resulting data with the example
    Metriq score reported in Table~\ref{tab:metriq-summary}.
    
    \item We present insight enabled by a shared benchmarking dataset by analyzing cross-benchmark structure and relationships between
    benchmark outcomes and public calibration metrics (Section~\ref{sec:cross-benchmark}).
\end{enumerate}

The remainder of this paper is organized as follows.
Section~\ref{sec:methods} describes the platform architecture,
followed by the Metriq score in Section~\ref{sec:metriq-score}.
Section~\ref{sec:benchmarking-suite} details the benchmark suite and its
implementation along with cross-platform hardware results, followed by the cost
estimation in Section~\ref{sec:cost-estimation}. We then analyze cross-benchmark
correlations in Section~\ref{sec:cross-benchmark}; and conclude in
Section~\ref{sec:discussion} with a discussion of limitations and future
directions.

The aim of this work is neither to claim a final benchmark authority, nor to present the
current suite or composite score as definitive. Rather, it is to put a working
platform into the open that researchers, labs, vendors, and standards groups can inspect, run, critique, and extend. In this sense, the
work is both a report on benchmarking methodologies that already exist and an open
invitation to collaborate on what quantum benchmarking should become.

\tcbset{
  colback=white,
  colframe=gray,
  boxrule=0.4pt,
  sharp corners,
  left=6pt,right=6pt,top=6pt,bottom=6pt,
  fonttitle=\bfseries,
  breakable
}
\newtcolorbox{headerbox}[1]{title={#1}}

\begin{table}[ht]
    \centering
    \footnotesize
    \setlength{\tabcolsep}{3pt}
    \begin{tabularx}{\linewidth}{l l X r r r r r r r r r M}
    \toprule
    \textbf{Vendor} &
    \textbf{Cloud} &
    \textbf{Device} &
    \textbf{Qubits} &
    \textbf{BSEQ} &
    \textbf{EPLG} &
    \makecell{\textbf{Mirror} \\ \textbf{Circuits}}&
    \textbf{CLOPS}$^{\ddagger}$ &
    \makecell{\textbf{QML} \\ \textbf{Kernel}} &
    \makecell{\textbf{LR-} \\ \textbf{QAOA}} &
    \textbf{WIT} &
    \textbf{QFT} &
    \makecell{\textbf{Metriq} \\ \textbf{score}} \\
    \midrule

    \multicolumn{4}{l}{\scriptsize\itshape Example benchmark weights (scale-based) } &
    \scriptsize 0.2069 &
    \scriptsize 0.1494 &
    \scriptsize 0.1703 &
    \scriptsize 0.2069 &
    \scriptsize 0.0733 &
    \scriptsize 0.1494 &
    \scriptsize 0.0145 &
    \scriptsize 0.0293 &
    \multicolumn{1}{c}{}
    \\
    \midrule

    IBM &
    IBMQ &
    \texttt{ibm\_boston} &
    156 &
    135.51 &
    338.40 &
    625.62 &
    104.31 &
    190.94 &
    173.61 &
    116.07 &
    145.98 &
    252.61
    \\
    
    Quantinuum &
    NEXUS &
    \texttt{quantinuum\_h2\_2} &
    56 &
    58.08 &
    181.53 &
    539.88 &
    \NA\;$^{\ddagger}$      &
    361.42 &
    89.12 &
    126.89 &
    523.15 & 
    188.05
    \\

    IBM &
    IBMQ &
    \texttt{ibm\_pittsburgh} &
    156 &
    132.03 &
    258.53 &
    267.71 &
    103.74 &
    149.32	 &
    168.27 &
    118.11 &
    126.69 & 
    174.51
    \\
    
    IBM &
    IBMQ &
    \texttt{ibm\_kingston} &
    156 &
    125.09 &
    149.27 &
    371.68 &
    104.83 &
    171.57 &
    161.68	 &
    110.49 &
    93.17 & 
    174.23
    \\
    
    IBM &
    IBMQ &
    \texttt{ibm\_marrakesh} &
    156 &
    129.43 &
    201.11  &
    249.28 &
    102.30 &
    130.57 &
    152.69 &
    113.20 &
    80.23 & 
    156.82
    \\

    IBM &
    IBMQ &
    \texttt{ibm\_fez} &
    156 &
    134.64 &
    37.30 &
    173.07 &
    106.72 &
    119.37 &
    131.88 &
    111.90 & 
    58.28 & 
    116.77
    \\

    \rowcolor{gray!15}
    IBM &
    IBMQ &
    \texttt{ibm\_torino} &
    133 &
    100.00 &
    100.00 &
    100.00 &
    100.00 &
    100.00 &
    100.00 &
    100.00 &
    100.00 &
    100.00
    \\

    IQM &
    AWS Braket &
    \texttt{iqm\_emerald} &
    54 &
    56.53 &
    12.75 &
    $15.94^{\dagger}$&
    \NA\;$^{\ddagger}$&
    $25.87^{\dagger}$ &
    22.65 &
    85.97 &
    31.43 & 
    23.76
     \\
    
    IQM &
    AWS Braket &
    \texttt{iqm\_garnet} &
    20 &
    30.20 &
    9.84 &
    $2.99^{\dagger}$&
    \NA\;$^{\ddagger}$&
    $23.88^{\dagger}$ &
    11.62	 &
    92.24 &
    44.13 & 
    14.34
    \\
    
    Rigetti &
    AWS Braket &
    \texttt{rigetti\_ankaa\_3} &
    82 &
    5.72 &
    8.65&
    $0.28^{\dagger}$ &
    \NA\;$^{\ddagger}$ &
    $12.84^{\dagger}$ &
    0.48 &
    59.90 &
    4.50 &
    4.54
    \\
    
    OriginQ &
    OriginQ &
    \texttt{wukong\_72} &
    72 &
    6.34 &
    0.00$^*$ &
    0.04\phantom{$^*$} &
    \NA\;$^{\ddagger}$ & 
    11.20\phantom{$^*$} &
    2.47  &
    56.91 & 
    1.53 &
    3.38
    \\
    \bottomrule
    \end{tabularx}
    \vspace{0.5em}
    \raggedright\footnotesize
    \caption{Benchmarking data taken using \mgym v0.4--v0.6 across multiple
    quantum devices and cloud platforms. The table includes normalized
    results for benchmark tasks 
    %(BSEQ, EPLG, Mirror Circuits, QML Kernel, LR-QAOA, WIT, and the QFT)
    from the Metriq suite. Task scores are normalized against a reference device (shaded gray), set to 100.00. The rightmost
    column includes the aggregated \textbf{Metriq score} (using the example benchmark weights, full definition and
    context in Section~\ref{sec:metriq-score}). We encourage readers to recompute the score using alternative weighting choices that better match their preferred emphasis. Entries marked with $^{\dagger}$ were collected using AWS Braket verbatim mode, corresponding to a restricted compilation setting; Entries marked with $^*$ indicate unsuccessful runs (see the corresponding benchmark sections for details); the CLOPS score $^{\ddagger}$ contributes a zero value to the composite score for devices that do not provide the needed timing information. 
    \\
    \\
    \textbf{Disclaimer}: Values in the table were collected at various times between March 2025 and March 2026 and they are intended for reference only; they should not be interpreted as a definitive or fully up-to-date ranking of device performance. 
 }
    \label{tab:metriq-summary}
\end{table}

%------------------------------------------------------------------------------%
\section{Methods}
\label{sec:methods}
%------------------------------------------------------------------------------%

%------------------------------------------------------------------------------%
\subsection{The Metriq platform}
\label{sec:metriq-platform}
%------------------------------------------------------------------------------%

The Metriq platform follows a modular, file-based design that separates benchmark execution, data storage, and visualization. This architecture avoids reliance on centralized databases, enabling reproducibility, offline data analysis, and transparent community contributions. Execution results are exchanged between components exclusively through version-controlled artifacts, ensuring that all analyses can be traced back to benchmark records.

Each component can be used independently, but the full power of the platform is
unlocked when they are used together. Together, these components address the
challenges outlined in Section~\ref{sec:introduction}: they make benchmark
definitions explicit and reusable, decouple execution from presentation while
preserving provenance, and enable community-driven, cross-platform evaluation.
In this section, we will describe the three components, including the \mgym
runner, the dataset \mdata and the website \mweb.

%------------------------------------------------------------------------------%
\subsection{Components}
\label{sec:components}
%------------------------------------------------------------------------------%

%------------------------------------------------------------------------------%
\subsubsection{Runner: \mgym}
\label{sec:mgym-runner}
%------------------------------------------------------------------------------%
The \mgym runner can be viewed at two levels: as a \emph{framework} and as a
\emph{suite of benchmarks}. As a framework, it supports specifying a benchmark of interest and executing it across
multiple supported backends. By exposing primitives for parameter validation,
circuit construction, backend abstraction, and result logging,
\mgym substantially shortens the path from a benchmark proposed in a research
paper to a runnable and shareable implementation. This also makes it
straightforward to iterate on benchmark definitions and parameter choices, since
running updated versions on different devices reduces to a small set of uniform
commands. As a suite, \mgym includes ready-to-use implementations of benchmarks
that are of broad interest to the community and provide meaningful insight about
the capabilities of quantum computing devices. These reference benchmarks are
cross-platform (i.e., capable of running on several vendor backends) and, being
open source, are transparent and open to inspection and improvement.

The core of \mgym is implemented in Python, reflecting the fact that much of
the quantum software ecosystem, including Qiskit~\cite{javadiabhari2024quantum}
and Cirq~\cite{cirq2024developers}, is built around Python as a common
interoperability layer for quantum SDKs and tools. The design of \mgym is guided
by the following principles:
\begin{itemize}
    \item \emph{Open}: \mgym has been open source since its inception and is
    developed entirely in a public code repository. This allows for external
    contributions and reuse in both academic and industry settings.
    
    \item \emph{Transparent}: All benchmark parameters are declared in explicit
    schema files, and the corresponding benchmark protocol implementations are
    available for review by the community.
    
    \item \emph{Cross-platform}: \mgym is designed to run the very same
    benchmark logic across multiple quantum hardware providers. This allows for
    comparisons across heterogeneous platforms without duplicating benchmark
    implementations.
    
    \item \emph{User-friendly}: A simple command-line interface supports
    dispatching, fetching, and viewing benchmark jobs, so that even operating
    sophisticated protocols reduces to a few commands. This lowers
    the barrier to adoption and makes it practical to iterate on benchmarks in
    response to new ideas or hardware capabilities.
\end{itemize}

The standard user workflow life cycle in \mgym is depicted in
Fig.~\ref{fig:metriq-workflow}. The inclusion of benchmarks in \mgym follows a
community-driven process: benchmarks are proposed through a Request for Comments
(RFC) in the form of a repository discussion. The feasibility of this is
evaluated, leading to an implementation, which is reviewed in the form of a code
pull request. The current suite reflects benchmarks that are insightful but
could be easily implemented, understood, and executed across multiple vendor
platforms without introducing implicit architectural biases. Benchmark
parameters such as shot counts, circuit depths, and other experimental settings
are currently determined by our team based on established practices in the
literature and cost evaluation. As the platform matures, we intend for parameter defaults and changes to follow the same public code-review process as benchmark inclusion, with proposals discussed in the repository and captured in versioned schemas to make consensus and provenance explicit. 

A key contribution of \mgym is its ability to execute identical benchmark protocols across diverse cloud APIs and quantum hardware architectures including superconducting circuits and trapped ions using a unified interface. This cross-platform capability is enabled by leveraging the \qbraid SDK~\cite{qbraid2025sdk}, which provides a common interface for circuit transpilation, job execution, and credential management across multiple quantum hardware providers. Because transpilation choices can bias cross-platform comparisons, Metriq treats the compiler as part of the benchmarked stack and interprets results in that context; compilation settings and their impact are
discussed in Section~\ref{sec:benchmarking-compilation}. We extend this interface with primitives tailored to benchmarking, including functions to extract QPU execution time for a given task and to retrieve the device connectivity graph. The \mgym framework is designed to be extensible, so that additional benchmark-specific primitives can be implemented and made available to benchmark developers.

Benchmark configuration in \mgym is handled through explicit, machine-readable parameter schemas that are treated as first-class objects of the framework. Each benchmark defines a JSON schema specifying all configurable parameters, their types, allowed ranges, and default values. At run time, user-provided configurations are validated against this schema. This mechanism ensures that benchmark runs are precisely reproducible, prevents silent configuration drift across hardware providers, and guarantees that results generated on heterogeneous devices remain directly comparable. By enforcing formal parameter schemas, \mgym introduces a level of configuration transparency and reproducibility that is often missing in existing quantum benchmarking frameworks. To illustrate this mechanism, the schema box in the following page %Listing~\ref{box:params-schema}
shows an example schema used to define the configuration space of an example benchmark.

\begin{tcblisting}
{
  listing only,
  title={Example of a benchmark parameter schema},
  label={box:params-schema},
  fonttitle=\bfseries,
  float=htb,
}
{
  "$id": "metriq-gym/qml_kernel.schema.json",
  "$schema": "https://json-schema.org/draft/2020-12/schema",  
  "title": "QML Kernel",
  "description": "QML Kernel benchmark schema definition",
  "type": "object",
  "properties": {
    "benchmark_name": {
      "type": "string",
      "const": "QML Kernel",
      "description": "Name of the benchmark. Must be 'QML Kernel' for this schema."
    },
    "num_qubits": {
      "type": "integer",
      "description": "Number of qubits used in the QML Kernel circuit(s).",
      "minimum": 2,
      "examples": [10]
    },
    "shots": {
      "type": "integer",
      "description": "Number of measurement shots (repetitions) for the QML Kernel benchmark.",
      "default": 1000,
      "minimum": 1,
      "examples": [1000]
    }
  },
  "required": ["benchmark_name", "num_qubits"]
}
\end{tcblisting}

The \mgym tool is released via the Python Package Index (PyPI). The
documentation includes a user guide, API reference, and tutorials. The \mgym
code is licensed under the permissive Apache-2.0 license to facilitate adoption
and integration within the evolving quantum software stack.

%------------------------------------------------------------------------------%
\subsubsection{Dataset: \mdata}
\label{sec:mgym-dataset}
%------------------------------------------------------------------------------%

The Metriq dataset is a version-controlled collection of structured benchmark records, stored as files in a GitHub repository and populated directly by the runner. To our knowledge, this is the first public dataset to systematically aggregate
benchmarking evidence across a heterogeneous set of quantum computers, spanning multiple providers, hardware architectures and devices, and benchmark types. The dataset is designed for continuous growth, allowing new results to be incorporated as they are collected.

The dataset is organized according to a simple and composable directory layout:
\begin{center}
    \verb|{source}/{version}/{provider}/{device}/{timestamp}_{benchmark-type}_{hash}.json|
\end{center}
Here, \texttt{source} identifies the origin of the data (\mgym or other tools),  \texttt{version} encodes the benchmark suite or protocol version,  \texttt{provider} and \texttt{device} specify the hardware backend, \texttt{timestamp} indicates the time of collection of the result,  and \texttt{benchmark-type} is the name of the benchmark protocol that was run.  This structure makes it straightforward to locate all results for a given device or provider, compare across protocol versions, and run automated aggregation or consistency checks over well-defined slices of the dataset. This file-system layout, together with schema-validated JSON results,  allows the dataset to be reproduced, versioned, and analyzed just by cloning the repository.

Although the present work focuses on data generated via \mgym, the dataset design is intentionally more general. The same schema and directory structure can accommodate benchmark results produced by other software frameworks, as well as curated data ingested from the literature. For example, results reported in
research publications or technical reports could be curated into the Metriq schema, enabling direct comparison between \mgym-generated experiments and previously published benchmarking results. Metriq data are released under the Creative Commons Attribution 4.0 International (CC-BY-4.0) license.

%------------------------------------------------------------------------------%
\subsubsection{Website: \mweb}
\label{sec:mgym-website}
%------------------------------------------------------------------------------%

The \mweb application is a lightweight, data-driven front end that
exposes the \mdata dataset as interactive time-series plots and tables. 
Users can toggle between ``Graph'' and ``Table'' views, filter by benchmark and provider, and download machine-readable bundles (Benchmarks JSON, Platforms index) to reproduce analyses offline; platform summaries provide quick entry points into device-level drill-downs (see Appendix~\ref{app:metriq_web} for additional detail). The application is open source and decoupled from
execution, consuming schema-validated artifacts produced by \mdata.

Looking ahead, the website is designed to evolve into a broader hub for
benchmarking and performance analysis in quantum computing. Future releases will
integrate state-of-the-art resource-estimation data, allowing visitors to track
experimental progress against theoretical expectations. We also plan to expose
a traceable pathway from the website back to the underlying raw data, enabling
transparent inspection of provenance, configuration choices, and device
metadata. Commenting and annotation features will allow users and researchers
to discuss benchmark results directly on the platform, fostering community
dialogue around reproducibility, methodology, and emerging trends. Together,
these capabilities aim to make \mweb a central venue for tracking,
interpreting, and debating benchmarking evidence in a rapidly evolving field.

Together, \mdata, \mgym, and \mweb constitute an example of digital research
assets developed in accordance with the FAIR (Findable, Accessible,
Interoperable, and Reusable) principles~\cite{wilkinson2016fair}.

\begin{table}[th]
    \centering
    \small
    \setlength{\tabcolsep}{6pt}
    \renewcommand{\arraystretch}{1.15}
    \begin{tabular}{@{} l l l l l @{}}
    \toprule
    \textbf{Component} & \textbf{Repository} & \textbf{Stack} & \textbf{License} & \textbf{Docs / link} \\
    \midrule
    Runner  & \ufrepo{metriq-gym} & Python  & Apache-2.0 & \hreftexttt{https://unitaryfoundation.github.io/metriq-gym/}{unitaryfoundation.github.io/metriq-gym/} \\
    Dataset & \ufrepo{metriq-data} & Data (JSON) & CC-BY-4.0  & \hreftexttt{https://unitaryfoundation.github.io/metriq-data/}{unitaryfoundation.github.io/metriq-data/} \\
    Display  & \ufrepo{metriq-web} & TypeScript (Vega) & Apache-2.0 & \hreftexttt{https://metriq.info}{metriq.info} \\
    \bottomrule
    \end{tabular}
    \caption{Metriq platform components and public artifacts. Repository links are hosted on GitHub under the Unitary Foundation organization at
    \href{https://github.com/unitaryfoundation}{\texttt{github.com/unitaryfoundation}}.}
    \label{tab:platform-components}
\end{table}

%------------------------------------------------------------------------------%
\subsection{Developer and user workflow}
%------------------------------------------------------------------------------%
\begin{figure}
    \centering
    \includegraphics[width=0.75\linewidth]{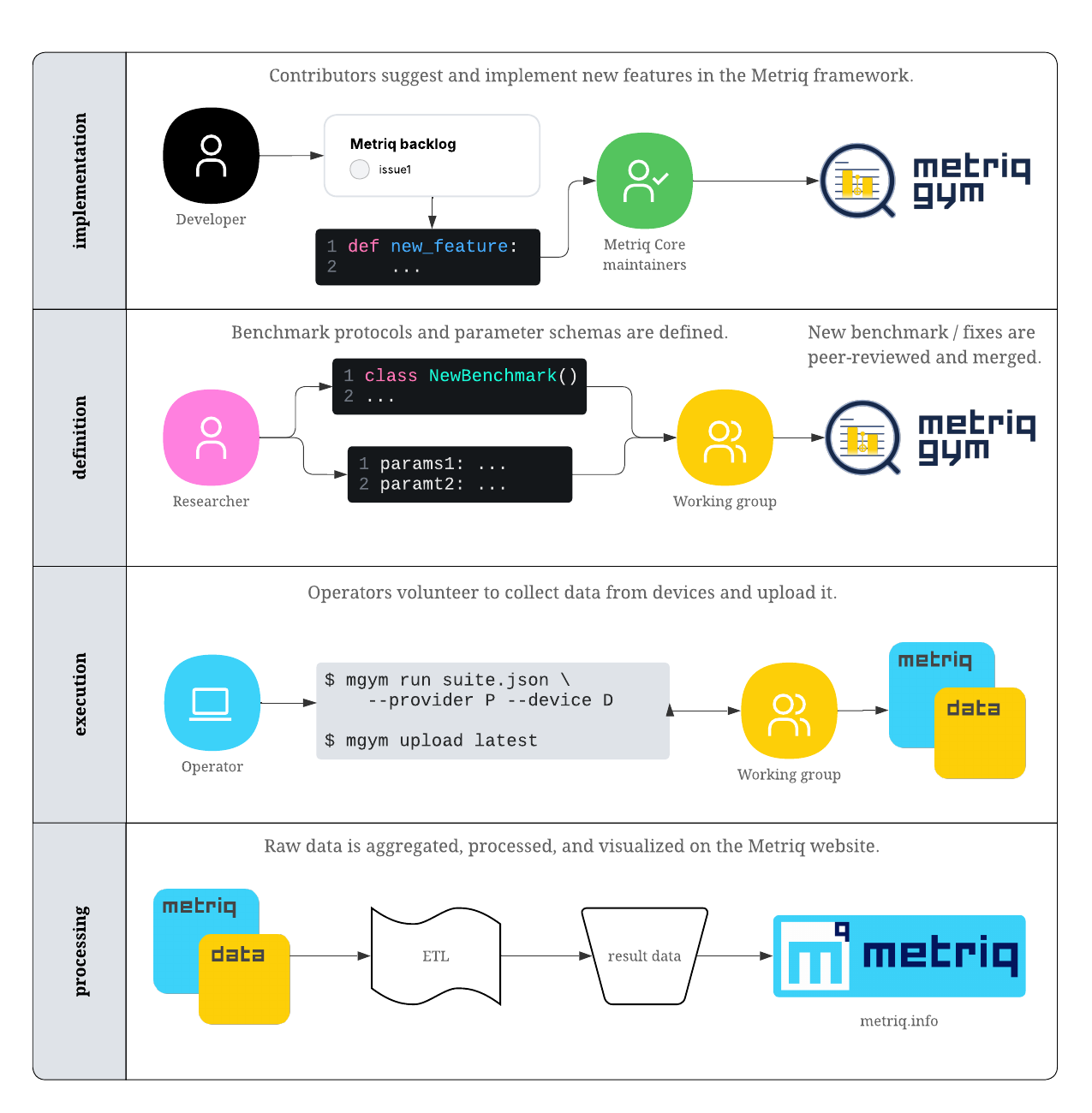}
    \caption{Developer and user workflow in Metriq. The workflow includes steps from feature implementation in \mgym, benchmark definition, to benchmark execution on devices or simulators, followed by data processing and publication.}
    \label{fig:metriq-workflow}
\end{figure}

The Metriq platform is designed to support a collaborative workflow that connects benchmark developers, operators, and the broader quantum ecosystem, as depicted in Fig.~\ref{fig:metriq-workflow}. At a high level, benchmark protocols are contributed and reviewed by the community, executed by operators across a wide range of quantum hardware, and ultimately
aggregated into the Metriq dataset and visualized on the Metriq website.

Developers begin by specifying the benchmark protocol and defining the associated parameter schema. These contributions are submitted through the public backlog, reviewed, and, once accepted, become part of the official benchmark suite available to all users.  Operators then execute the benchmarks by invoking the \mgym command-line interface (CLI). A key feature of the \mgym CLI is that it is \emph{resource-based}. Commands are structured as
\begin{center}
    \verb|mgym <resource> <action>|    
\end{center}
Here, the resource can be:
\begin{itemize}
    \item \verb|job|: a single benchmark instance, that is, one protocol with a
    specific set of parameters executed once on a backend.

    \item \verb|suite|: a curated collection of benchmark instances that are
    dispatched together. Running a suite ensures that all constituent benchmarks
    are evaluated under the same backend version, calibration epoch, and queuing
    conditions, providing a coherent snapshot of device performance.
\end{itemize}

A typical workflow might begin with:
\begin{center}
    \verb|mgym suite dispatch uf_complete.json --provider ibm --device ibm_torino|
\end{center}
which submits a suite of benchmarks for dispatching to a single backend. After completion, results are fetched via the \verb|mgym poll| action,  and finally uploaded using \verb|mgym upload|, contributing raw benchmark evidence to the shared dataset.

A defining feature of \mgym is its \emph{asynchronous dispatch/poll execution model}. Instead of requiring operators to wait synchronously for hardware jobs to finish, which often includes substantial queueing time, \mgym dispatches tasks to the provider and immediately returns control to the user. A lightweight polling mechanism then monitors job status and retrieves completed results in the background. This architecture is essential for practical large-scale or multi-backend benchmark campaigns and enables automation from laptops, batch schedulers, and CI systems without interactive sessions. 
It also avoids tying the user experience to unpredictable hardware queue times, 
a common bottleneck in traditional approaches.

This workflow applies equally to cloud-accessible hardware and to local
experimental platforms. Research groups operating their own quantum
devices such as university laboratories are encouraged to run \mgym directly
on their in-house systems, allowing their devices to be benchmarked using the
same protocols, schemas, and reproducibility guarantees as cloud platforms.

Once uploaded results are approved and merged into the dataset, the platform's
ETL pipeline aggregates, validates, and processes these data for presentation on
the Metriq website. This end-to-end workflow is summarized in
Fig.~\ref{fig:metriq-workflow}, which illustrates the interactions between
developers, operators, and the Metriq platform.

%------------------------------------------------------------------------------%
\subsubsection{Provider integration}
\label{sec:metriq-gym-providers}
%------------------------------------------------------------------------------%

A distinctive feature of \mgym is its broad and modular coverage of quantum hardware and cloud access pathways. Because benchmark protocols are expressed independently of the execution backend, \mgym can target the full cross-product of supported cloud providers and hardware vendors, enabling different platforms to be benchmarked through a single unified workflow. Our benchmarking framework is among the first attempts of spanning this range of provider combinations.
The cloud and hardware configurations supported in the current release are listed in Table~\ref{tab:cloud-hw}, but the architecture is intentionally extensible: developers can integrate new providers by implementing a small, clearly defined interface layer. Importantly, \mgym is not limited to commercial cloud access.

\begin{table}[H]
    \centering
    \begin{tabular}{lcccccc}
        \multirow{2}{*}{\textbf{Cloud} $\downarrow$}
          & \multicolumn{6}{c}{\textbf{Hardware} $\rightarrow$} \\
        \cmidrule(lr){2-7}
         & \textbf{IBM} & \textbf{IonQ} & \textbf{IQM} & \textbf{OriginQ} & \textbf{Quantinuum} & \textbf{Rigetti} \\
        \midrule
        AWS Braket        & -- & \cmark & \cmark & -- & -- & \cmark \\
        Azure Quantum     & -- & \cmark & -- & -- & \cmark & \cmark \\
        IBM Quantum       & \cmark & -- & -- & -- & -- & -- \\
        IonQ Quantum      & -- & \cmark & -- & -- & -- & -- \\
        Origin Quantum    & -- & -- & -- & \cmark  & -- & -- \\
        Quantinuum Nexus  & -- & -- & -- & -- & \cmark & -- \\
        \bottomrule
    \end{tabular}
    \caption{Cloud providers supported in \mgym{} v0.7 (platform-integrated offerings). Entries with ``--'' indicate combinations not provided through the listed platform's integrated access path and execution software stack (even if reachable via passthrough routes such as Quantinuum Nexus \emph{Linked Accounts}), rather than unsupported cases within \mgym{}. Aggregator platforms (e.g., qBraid Cloud) are not listed, to keep the table focused on platform-integrated access paths and their associated execution software stacks.}
    \label{tab:cloud-hw}
\end{table}
Despite the above provider coverage, integrating heterogeneous backends remains technically challenging. A central difficulty is the lack of uniform exposure of device and job metadata across providers: some platforms report calibration versions, qubit connectivity, or compiler revisions, whereas others omit this information entirely. Speed-oriented benchmarks pose a related
challenge, as they require access to the \emph{QPU execution time} of each job, excluding queueing delays and wire latency; only a subset of providers expose a reliable timestamp or duration field that can be interpreted in this way. Even
fundamental quantities such as qubit topology, gate durations, error rates, or the mapping of logical qubits to physical indices may be reported in provider-specific formats or not at all. These inconsistencies motivate the backend abstraction layer in \mgym, which isolates provider-specific details
behind a standard interface and surfaces only schema-validated metadata to the
benchmarking pipeline. As the landscape evolves, we expect continued progress
towards richer and more standardized metadata, which will in turn enable more
precise and comparable benchmark execution across devices.

%------------------------------------------------------------------------------%
\subsubsection{Local simulators}
\label{sec:simulators}
%------------------------------------------------------------------------------%

Beyond cloud execution, the \mgym runner supports seamless execution on local
simulators running on the operator's classical machine. In our design, local
simulators are treated as first-class devices within the same provider--device
hierarchy as hardware backends, using a provider label that we introduce,
\texttt{provider = local}. The backend abstraction makes it straightforward to
run benchmarks against a noiseless state-vector simulator (e.g.
\texttt{qiskit-aer}) or noise-instrumented ``fake'' devices (e.g. IBM's
\texttt{fake\_torino}.) This enables developers to validate benchmark schema and
protocols, allowing for rapid iteration, before spending any cloud device quota.
Because simulators participate fully in the metadata, schema, and provenance
machinery of \mgym, their outputs remain directly comparable to hardware runs.
For example, a typical invocation such as

\begin{center}
    \verb|mgym job dispatch mc_config.json --provider local --device aer_simulator|
\end{center}
executes a Mirror Circuit benchmark on a local Qiskit Aer simulator using the
same workflow as for a cloud backend. This symmetry makes it straightforward to
prototype benchmarks on simulators and then migrate identical configurations to
hardware, or to use simulators as stable anchors when tracking device drift over
time.

The backend layer is modular, so that adding support for additional local
simulators requires only a thin adapter that implements hooks to execute a
quantum job and retrieve simulator metadata. We encourage developers to
contribute integrations for other widely used simulators. 

%------------------------------------------------------------------------------%
\subsubsection{Integration with existing benchmark software}
\label{sec:integration-with-existing}
%------------------------------------------------------------------------------%
All benchmarks in \mgym follow a common implementation pattern designed to standardize new contributions and simplify integration with existing benchmark frameworks. In accordance with the runner's asynchronous execution model, each benchmark provides a dispatch and poll handler as defined in the base \textit{Benchmark} class. This class serves as the primary template and directs developers to the core components of a benchmark: \textit{BenchmarkData}, \textit{BenchmarkResult}, and \textit{BenchmarkScore}. 

The dispatch handler returns a \textit{BenchmarkData} object that records benchmark-specific metadata (e.g., number of qubits and circuit parameters), which can then be accessed by the poll handler. After execution completes, the poll handler analyzes the results and produces a \textit{BenchmarkResult} containing a \textit{BenchmarkScore}. Because each of these components is modular and clearly scoped, they can be cleanly connected to external tooling, enabling existing benchmark suites to be incorporated into \mgym with minimal friction. 

An example of this is an integration with the QED-C Application-Oriented Benchmarks for Quantum Computing suite~\cite{lubinski2023application}.  The
QED-C benchmark suite provides well-defined, algorithmically diverse, and application-inspired workloads with standardized analysis procedures. Like \mgym, the QED-C benchmark suite decouples problem generation, circuit execution, and results analysis~\cite{patel2025platform}. However, the QED-C
suite lacks the consolidated provider integration and job management provided by \mgym. By wrapping selected QED-C benchmark modules, the \mgym codebase exposes QED-C benchmarks via the same standardized schema files used for natively
defined benchmarks. Internally, \mgym then invokes the QED-C code for generating circuits and analyzing results. This enables users to tap into the suite of QED-C benchmarks alongside the strength of \mgym's operational workflows.

%------------------------------------------------------------------------------%
\section{Metriq score}
\label{sec:metriq-score}
%------------------------------------------------------------------------------%

To summarize performance across the entire suite of heterogeneous benchmarks, we
introduce the \emph{Metriq score} (MS), a composite indicator that assigns one
scalar value to each device. The Metriq scores reported in
Table~\ref{tab:metriq-summary} are computed using the definition from this
section, applied to the set of results available for each device at the time of
evaluation.

Unless specified otherwise, in general, the construction proceeds in three steps:
(i) \emph{within-benchmark} aggregation across circuit widths,
(ii) baseline normalization of the resulting benchmark-level aggregates, and
(iii) \emph{across-benchmark} aggregation using an explicitly specified weighting
scheme.

Let $\mathcal{C}=\{1,\dots,K\}$ index the recorded results (\emph{components}) of
a given suite version (a \emph{series}) $s$. Each component $i\in\mathcal{C}$ is
specified by a benchmark label $b_i$, a metric $m_i$ within that benchmark, and
a selector $\sigma_i$ that fixes the relevant run parameters (e.g.\ ``QML Kernel
accuracy at 10 qubits''). We write $n_i\in\mathbb{N}$ for the circuit width
(number of qubits) associated with $\sigma_i$.
For a device $d$ and series $s$, denote by $v_i(d,s)$ the measured value for the
most recent result matching $(b_i,m_i,\sigma_i)$, and by
$v_i(d^{\mathrm{base}},s)$ the corresponding value for the designated baseline
device $d^{\mathrm{base}}$.

In the suite versions considered here, each benchmark contributes a single
headline metric (so the metric is implicitly determined by $b$). We therefore
group components by benchmark label. Let
\begin{equation}
  \mathcal{B} := \{\, b_i : i\in\mathcal{C} \,\},
  \qquad
  \mathcal{C}_b := \{\, i\in\mathcal{C} : b_i=b \,\} \ \ (b\in\mathcal{B}).
\end{equation}

\textit{Step (i): linear aggregation across circuit width (within each benchmark).}
Many benchmarks are evaluated at multiple circuit widths. To obtain a single
raw benchmark value per device, we form a width-weighted linear aggregate. For
each component $i\in\mathcal{C}$ we define the \emph{within-benchmark} weight
\begin{equation}
  \alpha_i
  := \frac{n_i}{\sum_{j\in\mathcal{C}_{b_i}} n_j},
  \qquad i\in\mathcal{C},
  \label{eq:width-weights}
\end{equation}
so that $\sum_{i\in\mathcal{C}_b}\alpha_i=1$ for every benchmark $b$.
The width-aggregated raw value for benchmark $b$ is then
\begin{equation}
  \bar{v}_b(d,s)
  := \sum_{i\in\mathcal{C}_b} \alpha_i\, v_i(d,s),
  \qquad b\in\mathcal{B}.
  \label{eq:width-aggregate}
\end{equation}
Because Eq.~\eqref{eq:width-aggregate} is linear, we apply this aggregation at
the level of the raw measured values before performing baseline normalization.
If the required measurements for benchmark $b$ are missing for device $d$ in
series $s$, we define the benchmark subscore below to be zero, thereby penalizing
missing coverage.

\textit{Step (ii): baseline normalization (at the benchmark level).}
We next normalize the width-aggregated benchmark values relative to the
baseline. If higher values indicate better performance for benchmark $b$, we set
\begin{equation}
  \mathrm{BS}_b(d,s)
  := 100 \cdot \dfrac{\bar{v}_b(d,s)}{\bar{v}_b(d^{\mathrm{base}},s)}.
  \label{eq:benchmark-normalization}
\end{equation}
For benchmarks where lower values are better (e.g.\ error rates), we invert the
ratio:
\begin{equation}
  \mathrm{BS}_b(d,s)
  := 100 \cdot \dfrac{\bar{v}_b(d^{\mathrm{base}},s)}{\bar{v}_b(d,s)}.
  \label{eq:benchmark-normalization-inverted}
\end{equation}
In both cases, a value of $100$ corresponds to parity with the baseline device,
values greater than $100$ indicate better-than-baseline performance by that
factor, and values below $100$ indicate worse performance. If benchmark $b$ is
missing for device $d$ in series $s$, we set $\mathrm{BS}_b(d,s):=0$.

\textit{Step (iii): across-benchmark aggregation with scale-aware benchmark weights.}
\label{para:weight-aggregation}
To combine heterogeneous benchmark subscores into a single scalar, we assign
each benchmark $b$ a non-negative weight $w_b$ with $\sum_{b\in\mathcal{B}}w_b=1$.
Because larger circuit widths probe regimes where classical simulation becomes increasingly difficult and where hardware errors accumulate more strongly, we weight benchmark contributions according to their effective circuit scale.
We define an \emph{effective width} for each benchmark as the width-weighted mean
of its circuit widths under Eq.~\eqref{eq:width-weights}:
\begin{equation}
  \mu_b
  := \sum_{i\in\mathcal{C}_b} \alpha_i\, n_i
  = \dfrac{\sum_{i\in\mathcal{C}_b} n_i^2}{\sum_{i\in\mathcal{C}_b} n_i},
  \qquad b\in\mathcal{B}.
  \label{eq:effective-width}
\end{equation}
For example, if a benchmark is evaluated at widths $\{10,20,50,100\}$, then
$\alpha=(0.056,\,0.111,\,0.278,\,0.556)$ and $\mu_b\approx 72.2$.

For benchmarks evaluated at a single width $n^\star$ (e.g. CLOPS and WIT), this reduces trivially to
$\mu_b=n^\star$. 
For whole-device benchmarks without a natural width sweep, such as BSEQ, we assign a declared reference scale $n_b^{\mathrm{ref}}$ as part of the suite specification; in the present series of the suite,
we take $n_{\mathrm{BSEQ}}^{\mathrm{ref}}=100$. 

We then normalize these effective widths over the suite to obtain benchmark
weights
\begin{equation}
  w_b
  := \dfrac{\mu_b}{\sum_{c\in\mathcal{B}}\mu_c},
  \qquad b\in\mathcal{B}.
  \label{eq:benchmark-weights}
\end{equation}
This default choice transparently emphasizes performance in larger-width regimes,
which are typically more challenging and more informative about scalable device
capability.

Finally, the Metriq score for a device $d$ in series $s$ is defined as the
weighted average of benchmark subscores,
\begin{equation}
  \mathrm{MS}(d,s)
  := \sum_{b\in\mathcal{B}} w_b \, \mathrm{BS}_b(d,s).
  \label{eq:metriq-score}
\end{equation}
Eq.~\eqref{eq:metriq-score} yields a leaderboard-style scalar value that
facilitates direct comparison across devices, while ensuring that the influence
of each benchmarked capability is explicitly encoded in the benchmark weights
$w_b$.

As an example, consider a configuration for a fictitious suite version
\texttt{v0.4}. Assume that the baseline device is \texttt{ibm\_torino}, and that
the suite contains $K=3$ components: one \texttt{BenchA} result at $n_1=56$ qubits,
and two \texttt{BenchB} results at $n_2=10$ and $n_3=20$ qubits.

Suppose that the corresponding measured values are
\begin{equation*}
      v_1(d^{\mathrm{base}})= 0.60,\  v_1(d) = 0.75; \quad
      v_2(d^{\mathrm{base}})= 0.82,\  v_2(d) = 0.88; \quad
      v_3(d^{\mathrm{base}})= 0.76,\  v_3(d) = 0.70.
\end{equation*}
Within the \texttt{BenchB} benchmark, Eq.~\eqref{eq:width-weights} gives
$\alpha_2=\tfrac{10}{10+20}=\tfrac{1}{3}$ and $\alpha_3=\tfrac{2}{3}$, hence
Eq.~\eqref{eq:width-aggregate} yields
$\bar v_{\mathrm{BenchB}}(d^{\mathrm{base}})=\tfrac{1}{3}\cdot 0.82+\tfrac{2}{3}\cdot 0.76=0.78$
and
$\bar v_{\mathrm{BenchB}}(d)=\tfrac{1}{3}\cdot 0.88+\tfrac{2}{3}\cdot 0.70=0.76$.
The benchmark subscore is then
$\mathrm{BS}_{\mathrm{BenchB}}(d,s)=100\cdot 0.76/0.78\approx 97.4$ via
Eq.~\eqref{eq:benchmark-normalization}. For \texttt{BenchA}, we have
$\bar v_{\mathrm{BenchA}}(d^{\mathrm{base}})=0.60$ and $\bar v_{\mathrm{BenchA}}(d)=0.75$,
so $\mathrm{BS}_{\mathrm{BenchA}}(d,s)=125$.

Eq.~\eqref{eq:effective-width} gives
$\mu_{\mathrm{BenchA}}=56$ and $\mu_{\mathrm{BenchB}}=(10^2+20^2)/(10+20)=16.7$, so
Eq.~\eqref{eq:benchmark-weights} yields benchmark weights
$w_{\mathrm{BenchA}}\approx 0.771$ and $w_{\mathrm{BenchB}}\approx 0.229$.
Finally, Eq.~\eqref{eq:metriq-score} gives
\begin{equation*}
  \mathrm{MS}(d,s)
  \approx 0.771\cdot 125 + 0.229\cdot 97.4
  \approx 118.7.
\end{equation*}
In this example, the resulting Metriq score of approximately $118.7$ indicates
that device $d$ outperforms the baseline by about $19\%$ under the specified
mixture of benchmarks, with larger-width evaluations receiving greater emphasis
through the weights in Eqs.~\eqref{eq:width-weights}--\eqref{eq:benchmark-weights}.
In practice, the Metriq score is computed automatically within the ETL stage of
the Metriq workflow (Fig.~\ref{fig:metriq-workflow}) whenever new benchmark
results are ingested.

The Metriq score should be interpreted as a composite index over heterogeneous benchmark outcomes.
It is not a physical fidelity, and not an estimator of a single latent hardware capability. 
Each benchmark contributes a normalized subscore derived from its own task-specific metric (e.g.,
polarization, success probability, approximation ratio, throughput), and the overall score aggregates
these dimensionless quantities under an explicit weighting scheme.

It is also worth remarking that the baseline device simply provides an interpretable reference point: $\mathrm{BS}_b(d^{\mathrm{base}},s)=100$ for each benchmark $b$ where the baseline is measured on the benchmark's designated width grid. Changing the baseline corresponds to benchmark-dependent rescaling of the subscores; since this
rescaling generally differ across benchmarks, the composite ranking under Eq.~\eqref{eq:metriq-score} can in principle change. We therefore treat the baseline as part of the series specification and report it explicitly. Next, the benchmark weights $w_b$ encode explicit value judgments about which
capabilities and scales should dominate the composite. In this work we adopt a scale-aware default: within each benchmark we weight instances proportionally to circuit width (Eq.~\eqref{eq:width-weights}), and across benchmarks we weight
benchmarks proportionally to their effective width (Eqs.~\eqref{eq:effective-width} and \eqref{eq:benchmark-weights}). Alternative weighting schemes can be substituted in future suite versions in a transparent and reproducible way.
Furthermore, the final Metriq score is a linear, convex combination of benchmark subscores (Eq.~\eqref{eq:metriq-score}). This preserves interpretability, avoids
opaque nonlinear interactions between heterogeneous metrics, and ensures that trade-offs arise only through explicitly stated weights rather than implicit modeling assumptions. The construction is conceptually aligned with classical
performance indicators such as Epoch AI's Epoch Capabilities Index for LLM models~\cite{epoch2025eci} and the Geekbench score for classical CPUs~\cite{primatelabs2025geekbench}, adapted here to the setting of quantum-device benchmarking.

While some benchmarks naturally yield uncertainty estimates (and we show these as error bars when available), others are currently summarized as single-point metrics. In this work, Metriq scores aggregate point estimates; propagating
uncertainty through the composite score is left to future work.

%------------------------------------------------------------------------------%
\section{Benchmarking suite}
\label{sec:benchmarking-suite}
%------------------------------------------------------------------------------%

\begin{table*}[t]
    \centering
    \small
    \setcellgapes{1.5pt}\makegapedcells
    \renewcommand\theadfont{\normalsize\bfseries}
    \begin{tabularx}{\textwidth}{l *{10}{c}}
    \toprule
    & \multicolumn{4}{c}{\textbf{Scalable}} 
    & \multicolumn{5}{c}{\textbf{Full-stack sensitivity}}
    & \multicolumn{1}{c}{\textbf{Practicality}}\\
    \cmidrule(lr){2-5}\cmidrule(lr){6-10}\cmidrule(lr){11-11}
    \textbf{Benchmark} 
    & \makecell{QPU-\\efficient\\(shots)}
    & \makecell{Hardware-\\native}
    & \makecell{Verifiable}
    & \makecell{Meaningful \\ scaling up}
    & \makecell{Cross-\\talk}
    & \makecell{SPAM}
    & \makecell{Transpilation\\pressure}
    & \makecell{Connectivity\\dependence}
    & \makecell{Parallel-\\ism}
    & \makecell{Application \\relevance}\\
    \midrule
    BSEQ & \strong & \strong & \strong & \strong & \notfocus & \strong & \notfocus & \strong & \strong & \notfocus \\
    CLOPS & \strong & \strong & \notfocus & \strong & \notfocus & \notfocus & \strong & \notfocus & \strong & \partialok \\
    EPLG & \strong & \strong & \strong & \strong & \strong & \notfocus & \notfocus & \notfocus & \strong & \notfocus \\
    LR-QAOA & \partialok & \strong & \strong & \strong & \strong & \strong & \strong & \strong & \partialok & \strong \\
    Mirror circuits & \strong & \strong & \strong & \strong & \partialok & \strong & \strong & \partialok & \strong & \notfocus \\
    QFT & \strong & \strong  & \strong & \strong & \strong & \strong &  \strong  & \strong & \partialok & \strong \\
    QML Kernel & \strong & \strong & \strong & \strong & \strong & \strong & \strong & \partialok & \strong & \strong    \\
    WIT & \strong & \partialok & \strong & \partialok & \strong & \partialok & \strong & \strong & \partialok & \strong \\
    \midrule
    \multicolumn{11}{l}{\strong~strong; \partialok~partial/conditional; \notfocus~not a primary focus.}
    \\
    \bottomrule
    \end{tabularx}
    \caption{Benchmarking methods in the \mgym benchmark suite. Table design inspired by~Gambetta \cite{gambetta2025charting}.}
    \label{tab:metriq_suite_matrix}
\end{table*}

We next discuss the suite of benchmarks in \mgym, where we customize and
standardize the implementation of existing benchmarks and also include the
design of new benchmark methods. The collection presented here constitutes a
first, intentionally opinionated version of what we consider to be a practical
and informative benchmark suite for near-term quantum devices. It reflects our
assessment of which metrics currently provide meaningful, reproducible insight
under realistic resource constraints, while being designed in a manner that
remains scalable as devices increase in size, connectivity, and operational
fidelity.

The benchmarks included were selected with two primary considerations. First, we
prioritized frugality: benchmarks that are monetarily inexpensive to run on
currently accessible quantum hardware. This ensures that the broader quantum
computing community, including academic researchers and smaller organizations,
can reproduce and validate our results without prohibitive costs. Second,
despite this constraint on resource consumption, we aimed to include benchmarks
that probe important device characteristics at scale, providing meaningful
insights into hardware capabilities across multiple performance dimensions and
remaining extensible to larger and more capable architectures. We emphasize that
this specific suite should not be interpreted as definitive or exhaustive.
Rather, it represents a starting point for structured and transparent
benchmarking within the Metriq platform. We expect the suite to evolve through
community input, empirical experience, and the inclusion of additional methods
spanning component-level, system-level, and application-inspired benchmarks in
future releases.

As shown in Table~\ref{tab:metriq_suite_matrix}, the suite is intentionally
heterogeneous, spanning system-level metrics that characterize fundamental
device properties (such as coherence and gate fidelity) as well as
application-inspired circuit benchmarks that capture structured circuit families
commonly encountered in quantum algorithms. This dual perspective enables
evaluation of both the intrinsic quality of quantum hardware and its practical
behavior under structured computational workloads. Inspired by
Ref.~\cite{gambetta2025charting}, Table~\ref{tab:metriq_suite_matrix}
characterizes each benchmark along the axes of scalability, full-stack
sensitivity, and practicality. The former includes whether the benchmark
requires few quantum samples (QPU-efficient), runs directly on native
gates/couplers (Hardware-native), has classical correctness check (Verifiable)
and has meaningful values when increasing circuit width or depth (Meaningful
scaling up). While for full-stack sensitivity, we consider if the selected
benchmark is sensitive to simultaneous inter-qubit errors (Crosstalk) and
state-preparation and measurement errors (SPAM), as well as whether its
performance depends strongly on mapping and decomposition to native gate set
(transpilation pressure) and whether it is constrained by coupling graph
(Connectivity dependence) and whether it benefits from many concurrent instances
(Parallelism).

In the subsections that follow, we provide detailed results for both system-level and application-inspired benchmarks currently supported in \mgym. Section~\ref{sec:benchmark-system} covers system-level metrics including BSEQ, EPLG, mirror circuits and CLOPS, with results from Quantinuum, IBM, Rigetti, IQM, and other supported providers. Section~\ref{sec:benchmark-application} presents application-inspired circuit benchmarks, such as workloads derived from
quantum machine learning kernel methods, which probe device performance on structured algorithmic tasks. All results discussed herein are publicly accessible via the \metriq platform. We encourage community members to explore
the underlying data, propose additional benchmarks protocols, and contribute to the ongoing development of this open benchmarking framework.

The benchmarks presented in this section define the components used in the composite Metriq score (Section~\ref{sec:metriq-score}), which we use for the summary comparison in Table~\ref{tab:metriq-summary}. For transparency, Appendix~\ref{app:benchmark-configs} includes a benchmark provenance table (Table~\ref{tab:benchmark-configs}) that records the configuration parameters, scale datapoints, and effective benchmark scales $\mu_b$ used in this work, making explicit how the present suite connects to the default benchmark-weight construction of Section~\ref{sec:metriq-score}.

%------------------------------------------------------------------------------%
\subsection{System-level benchmarks}
\label{sec:benchmark-system}
%------------------------------------------------------------------------------%

System-level benchmarks in \mgym target fundamental device characteristics that
are largely independent of specific algorithmic workloads. These benchmarks probe intrinsic hardware properties such as coherence, gate fidelity, connectivity, and circuit execution speed. Within the Metriq framework, they serve as a hardware-oriented baseline against which more structured, application-inspired circuit benchmarks can be interpreted.

%------------------------------------------------------------------------------%
\subsubsection{Bell state effective qubits (BSEQ)}
\label{sec:benchmark-bseq}
%------------------------------------------------------------------------------%
The Bell State Effective Qubits (BSEQ) benchmark evaluates a quantum device's
ability to produce high-quality entangled states across its entire connectivity
graph by testing for violations of the CHSH (Clauser-Horne-Shimony-Holt)
inequality~\cite{clauser1969proposed}. The CHSH inequality provides a
fundamental test of quantum nonlocality: for any pair of qubits, classical local
hidden variable theories predict that a certain combination of correlation
measurements cannot exceed 2, whereas quantum mechanics allows values up to
$2\sqrt{2} \approx 2.83$ for maximally entangled states~\cite{brunner2014bell}.

The benchmark proceeds as follows. First, the device's connectivity graph (coupling map) is analyzed and partitioned using edge coloring, which assigns colors to edges such that no two edges sharing a common vertex receive the same
color. This partitioning is critical because it identifies sets of qubit pairs that can be independently tested without interference: all pairs with the same
color can be measured simultaneously. For bipartite graphs, the optimal bipartite edge coloring algorithm is employed; otherwise, a greedy edge coloring heuristic is used.

\begin{figure}
    \centering
    \includegraphics[width=1\linewidth]{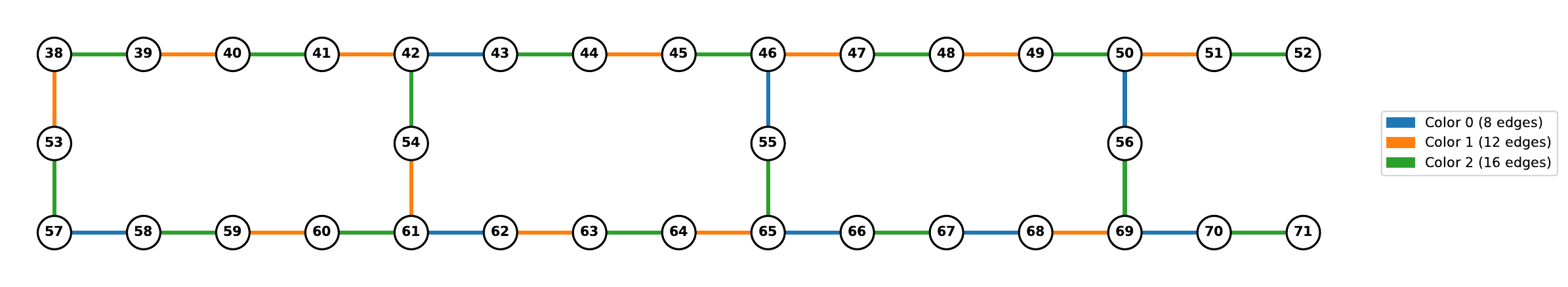}
    \caption{
      Edge coloring for the BSEQ benchmark on a representative section of the
      heavy-hex lattice topology found on IBM devices. Edges with the same color
      can be measured simultaneously, as they form an independent set where no
      two edges share a common qubit. The coloring partitions the couplings into
      3 color classes, reducing the required circuits from 144 (36 edges
      $\times$ 4 bases) to 12 (3 colors $\times$ 4 bases).
    }
    \label{fig:bseq_coloring}
\end{figure}

For each edge (qubit pair) in the device, the benchmark constructs a Bell state
by applying a Hadamard gate to the first qubit followed by a CNOT gate between
the pair, then applies a CHSH-optimal rotation $R_y(\pi/4)$ to the first qubit.
To test the CHSH inequality, four different measurement circuits are generated
for each edge, corresponding to the four correlation terms in the CHSH
expression: $\langle ZZ \rangle$, $\langle ZX \rangle$, $\langle XZ \rangle$,
and $\langle XX \rangle$, where $X$ and $Z$ denote Pauli measurement bases. The
appropriate basis rotations are applied before measurement, and the expectation
values are computed from the measurement statistics. The CHSH parameter $S$ is
computed as 
\begin{equation}
    S = |\langle ZZ \rangle + \langle ZX \rangle + \langle XZ \rangle - \langle XX \rangle|. 
\end{equation}
A qubit pair is considered to have violated the CHSH inequality if $S > 2$,
indicating non-classical correlations. The benchmark constructs a subgraph
containing only the edges (qubit pairs) that violate the inequality, and
computes two primary metrics: the \emph{largest connected component size}
(LCCS), which is the number of qubits in the largest connected component of this
subgraph, and the \emph{connection fraction}, defined as LCCS divided by the
total number of qubits in the device. The benchmark configuration is minimal:
\begin{verbatim}
"benchmark_name": "BSEQ",
"shots": 1000
\end{verbatim}
The \texttt{shots} parameter specifies the number of measurement repetitions per
circuit used to estimate correlation expectation values. We typically use $1000$
shots per circuit, balancing statistical precision against execution cost.
Unlike some benchmarks that require explicit circuit width or depth
specifications, BSEQ automatically adapts to the device's native connectivity by
testing all available qubit pairs, making it particularly suitable for comparing
devices with different architectures. For all-to-all devices, testing all qubit
pairs is costly without much added benefit given the density of the connectivity
graph. Instead, users can select a maximum number of colors to consider, which
reduces the number of qubit pairs tested and serves as a lower bound on the
LCCS.

The BSEQ benchmark characterizes device-wide entanglement capabilities, which
are essential for quantum algorithms that require multi-qubit entanglement, such
as variational quantum eigensolvers, quantum approximate optimization
algorithms, and quantum simulation protocols. Additionally, the use of the CHSH
inequality provides a theoretically grounded certification of genuine quantum
behavior, distinguishing it from operational benchmarks that may not rule out
classical explanations. 
The LCCS metric directly quantifies the scale at which a
device can maintain high-quality entanglement, a useful parameter for assessing
scalability toward practical quantum advantage. Finally, the connection fraction
metric provides insight into the uniformity of device performance: a high
connection fraction indicates that most qubits can participate in high-fidelity
entangled operations, whereas a low fraction suggests localized defects or
calibration issues.

Let $n$ denote the total number of qubits in the device and $\text{LCCS}$ the
size of the largest connected component of qubits successfully violating the
CHSH inequality. The connection fraction is simply
\begin{equation}
    f_{\text{conn}} = \frac{\text{LCCS}}{n}.
\end{equation}
This ratio lies in $[0,1]$, and a connection fraction near $1$ indicates that
nearly all qubits in the device can participate in high-fidelity entangled
operations, a key requirement for running large-scale quantum algorithms.
Devices with $f_{\text{conn}} < 0.5$ suggest either limited connectivity,
spatially correlated errors, or calibration issues affecting significant
portions of the processor. 

To compute an aggregate BSEQ score for inclusion in the overall Metriq score
(Section~\ref{sec:metriq-score}), we normalize both metrics against a baseline
device $d^{\mathrm{base}}$ and compute a weighted average. For a device $d$, the
normalized LCCS and connection fraction components are
\begin{equation}
    S_{\text{LCCS}}(d) = 100 \frac{\text{LCCS}(d)}{\text{LCCS}(d^{\mathrm{base}})}
    \quad \text{and} \quad
    S_{f_{\text{conn}}}(d) = 100 \frac{f_{\text{conn}}(d)}{f_{\text{conn}}(d^{\mathrm{base}})}.
\end{equation}
The BSEQ score is then defined as
\begin{equation}
\label{eq:bseq_definition}
    \mathrm{BSEQ~Score}(d) = \frac{7}{8} S_{\text{LCCS}}(d) + \frac{1}{8} S_{f_{\text{conn}}}(d).
\end{equation}
This weighting emphasizes absolute connectivity over the relative connection
fraction, reflecting the property that larger connected components enable
execution of broader algorithmic primitives, while the fraction primarily
indicates uniformity across the device. A BSEQ score of 100 indicates
performance equal to the baseline device. Table~\ref{tab:bseq-scores} summarizes
BSEQ metrics across all tested devices.

\begin{table}[H]
    \centering
    \begin{tabular}{lcccr}
        \toprule
        \textbf{Device} & \textbf{Qubits} & \textbf{LCCS} & \textbf{$f_{\text{conn}}$} & \textbf{BSEQ Score} \\
        \midrule
        \texttt{ibm\_boston} & 156 & 156 & 1.00 & 135.51 \\
        \texttt{ibm\_fez} & 156 & 155 & 0.99 & 134.64 \\
        \texttt{ibm\_pittsburgh} & 156 & 152 & 0.97 & 132.03 \\
        \texttt{ibm\_marrakesh} & 156 & 149 & 0.96 & 129.43 \\
        \texttt{ibm\_kingston} & 156 & 144 & 0.92 & 125.09 \\
        \texttt{ibm\_brisbane} & 127 & 120 & 0.94 & 106.74 \\
        \texttt{ibm\_torino} & 133 & 113 & 0.85 & 100.00 \\
        \texttt{quantinuum\_h2\_2} & 56 & 56 & 1.00 & 58.08 \\
        \texttt{iqm\_emerald} & 54 & 54 & 1.00 & 56.53 \\
        \texttt{iqm\_garnet} & 20 & 20 & 1.00 & 30.20 \\
        \texttt{wukong\_72} & 72 & 6 & 0.08 & 6.34 \\
        \texttt{rigetti\_ankaa\_3} & 82 & 6 & 0.07 & 5.72 \\
        \texttt{wukong\_102} & 102 & 1 & 0.01 & 0.77 \\
        \bottomrule
    \end{tabular}
    \caption{BSEQ benchmark metrics across tested devices. Qubits denote the
    total device qubit count. LCCS is the largest connected component size
    (number of qubits successfully demonstrating CHSH violations).
    $f_{\text{conn}}$ is the connection fraction (LCCS divided by total qubit
    count). The BSEQ Score is normalized against the baseline device, set to
    100.00; higher values indicate better performance. Results obtained with
    $1000$ shots between August -- December 2025.}
    \label{tab:bseq-scores}
\end{table}

%------------------------------------------------------------------------------%
\subsubsection{Error per layered gate (EPLG)}
\label{sec:benchmark-eplg}
%------------------------------------------------------------------------------%
Error Per Layered Gate (EPLG) is a recently introduced benchmark designed to
quantify how accurately a quantum processor can implement layers of two‑qubit
entangling gates across a connected set of qubits.  Whereas quantum volume
benchmarks focus on all-to-all circuits, the EPLG protocol is fine-tuned for
connectivity-constrained architectures. The protocol uses a subset of pairs of
qubits along a 1D chain, which evaluates the typical performance of devices with
nearest-neighbor connectivity, removing the overhead of routing and SWAPs
associated with executing circuits with arbitrary connectivity. The protocol was
first described in Ref.~\cite{mckay2023benchmarking} in 2023 as a scalable
alternative to quantum volume tests; in contrast to discrete pass/fail criteria,
EPLG produces a continuous figure of merit, and it is also not limited by
classical computation of the circuits. EPLG can be used to extract lower bounds
on two‑qubit error rates~\cite{mckay2023benchmarking}.

Consider a chain of $N$ connected qubits $\{q_0,\ldots,q_{N-1}\}$ on a device.
A layer of depth one consists of simultaneously applied two‑qubit gates $U_{ij}$
on disjoint pairs $(i,j)$ together with random single‑qubit Clifford gates on
idle qubits.  For the qubits that are connected only by nearest–neighbor
couplings, two disjoint sets of two‑qubit gates are needed to cover a linear
chain; more complex connectivity patterns may require additional layers.  Let
the disjoint layers be indexed by $m=1,\ldots,M$ and label the individual
elements within each layer by $j$; each element may be a two‑qubit gate $(i,j)$
or a single‑qubit idle.  A simultaneous direct randomized benchmarking
experiment is performed on each layer to estimate the process fidelity $F_{j,m}$
of each constituent subsystem.  The layer fidelity (LF) of disjoint layer~$m$ is
defined as the product of its process fidelities $\mathrm{LF}_m= \prod_{j}
F_{j,m}$, and the full layer fidelity is the product over disjoint
layers~\cite{mckay2023benchmarking}
\begin{equation}
  \label{eq:lf}
  \mathrm{LF} = \prod_{m=1}^{M} \mathrm{LF}_m.
\end{equation}
In the ideal case, without crosstalk, one can combine disjoint layer LFs multiplicatively to get the full LF; when crosstalk is
present they yield a lower bound on the layer fidelity~\cite{mckay2023benchmarking}.
Because $\mathrm{LF}$ decreases exponentially with the number of two‑qubit gates
$n_{2\mathrm{Q}}$ in the layer (for a linear chain $n_{2\mathrm{Q}}=N-1$), it is
normalized to a size‑independent error per layered gate defined by
\begin{equation}
  \label{eq:eplg-def}
  \mathrm{EPLG} = 1 - \mathrm{LF}^{1/n_{2\mathrm{Q}}}.
\end{equation}
This quantity thus characterizes the quality of the underlying native two‑qubit operations in their natural layered context.  For example, Ref.~\cite{mckay2023benchmarking} reported $\mathrm{EPLG}\approx1.7\times10^{-2}$ for a 127‑qubit
IBM Eagle device and $\mathrm{EPLG}\approx6.2\times10^{-3}$ for a 133‑qubit IBM Heron
device~\cite{mckay2023benchmarking}.

The protocol goes as follows: select a connected chain of $N$ qubits and decompose the parallel two-qubit layer into disjoint sublayers; 
run simultaneous direct RB with barriers inserted between sublayers to enforce synchronized execution
across a set of depths $l\in\texttt{lengths}$, fit decay rates to obtain process fidelities on active and idle subsystems using a three-parameter RB model $F(l)=a \alpha^l + b$ where $a$ and $b$ are fit parameters and $\alpha$ is the RB decay parameter for the active (or idle) subsystem, compute $\mathrm{LF}_m$ and then $\mathrm{LF}$, and finally get $\mathrm{EPLG}$.
Our implementation of EPLG follows the protocol above but introduces two
practical features to make the benchmark suitable for routine use in the
\mgym benchmark suite.  First, instead of exhaustively searching
for the optimal qubit chain~\cite{palacio2025parameter}, we sample a single simple path of length
$N$ from the device's coupling graph using a randomized algorithm.  The algorithm starts from a random edge
and extends the chain by randomly appending unused adjacent qubits, ensuring
that each selected edge supports the specified two‑qubit gate.  This random
chain approximates the typical performance of a chain of length $N$ without
requiring an expensive global search and pre-characterizations.  Second, the benchmark input allows the
user to specify the two‑qubit gate and the one‑qubit basis gates used to synthesize random Clifford operations. 

Adding EPLG to our benchmark suite provides complementary insight to other
metrics.  Whereas quantum volume and cross‑entropy benchmarking emphasize
algorithmic performance on short random circuits, EPLG directly targets the
error rates of native entangling operations executed in parallel across a large
region of the processor. Moreover, by using a random chain instead of the best
chain, our implementation yields a conservative estimate of device performance
that captures typical noise and crosstalk across the chip.  Including EPLG
alongside other benchmarks thus provides a more complete picture of hardware
quality, speed, and scale.  

The schema file includes \verb|num_qubits_in_chain| that defines the length $N$ of the
random chain, \verb|lengths| which is the list of RB depths $\ell$ for which
layered circuits are generated, \verb|num_samples| that specifies how many
random RB sequences are averaged, \verb|nshots| that controls the number of
shots per circuit.  The output of the benchmark is the estimated EPLG at chain
length $N$, along with intermediate results such as the layer fidelities at each
depth and the process fidelities of individual subsystems.

\begin{figure}
    \centering
    \includegraphics[width=0.8\linewidth]{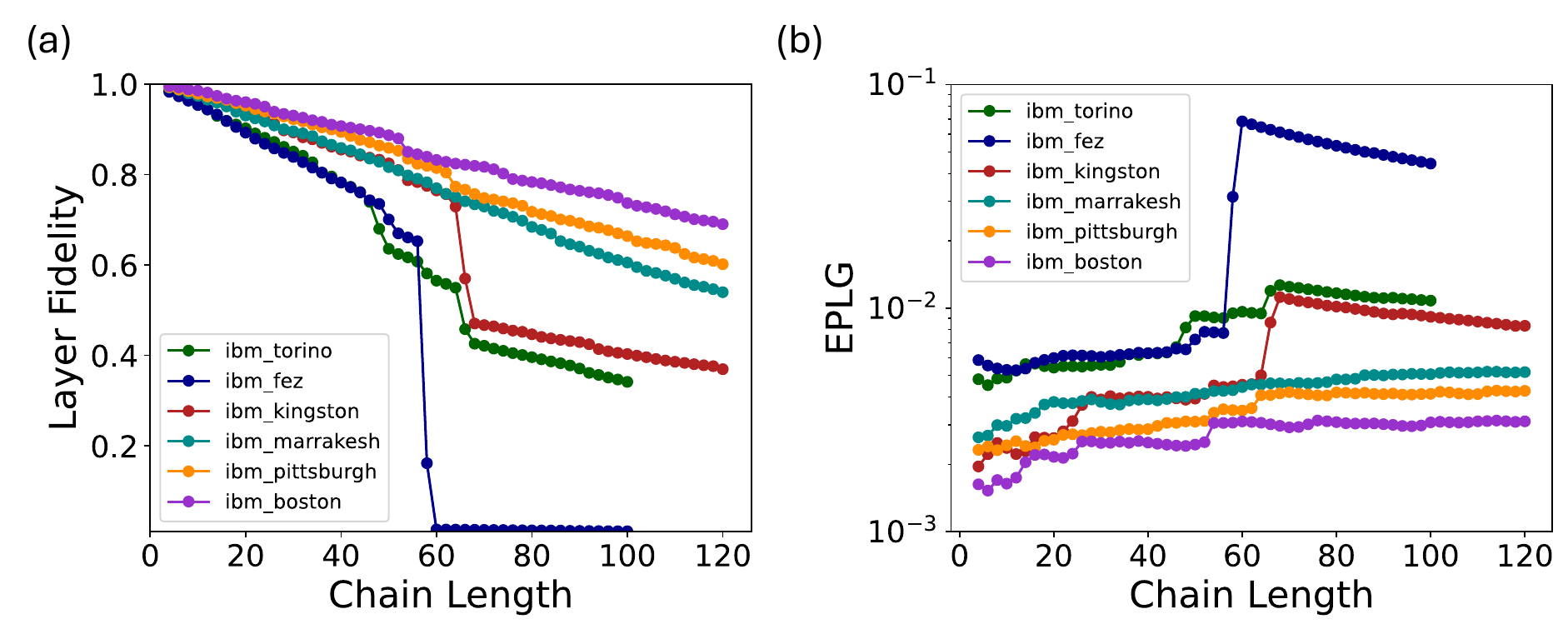}
      \caption{(a) Layer fidelity and (b) EPLG versus chain length on several selected IBM hardware. The list of RB layers is \texttt{lengths}= $[2,4,8,16,30,50,70,100,150,200,300,500]$ and we use 10 random RB instances per depth and 1000 shots per-circuit. }
    \label{fig:eplg_results}
\end{figure}

We next present the hardware results for EPLG benchmarking. Note that we obtain
low-qubit number EPLG values by post-processing the run at a large chain length.
For example, for devices with fewer than 100 qubits, we obtain EPLG-10 and
EPLG-20 by selecting length-10/20 subchains from the sampled $N=50$ path. The
results were executed on several quantum processors using the following
configuration for IBM and OriginQ devices:
\begin{verbatim}
"benchmark_name": "EPLG",
"num_samples": 10,
"shots": 1000,
"lengths": [2, 4, 8, 16, 30, 50, 70, 100, 150, 200, 300, 500]
\end{verbatim}
For IQM and Rigetti devices, we used a reduced configuration due to cost constraints:
\begin{verbatim}
"benchmark_name": "EPLG",
"num_samples": 5,
"shots": 500,
"lengths": [2, 4, 8, 16, 50, 100, 200, 300]
\end{verbatim}
% For \verb|quantinuum_h2_2| device, the following configuration was run due to cost constraints:
For the \texttt{quantinuum\_h2\_2} device, we used a reduced configuration due to cost constraints:
\begin{verbatim}
"benchmark_name": "EPLG",
"num_samples": 1,
"shots": 200,
"lengths": [2, 4, 8, 16, 30, 50, 70, 100, 150, 200]
\end{verbatim}

As an example, Fig.~\ref{fig:eplg_results} illustrates the measured
layer fidelity and corresponding EPLG versus chain length for IBM devices.  As the chain length
increases, the layer fidelity gradually declines on all devices due to the
accumulation of gate and readout errors.  The \verb|ibm_fez| and \verb|ibm_kingston|
devices show a precipitous drop in the layer fidelity.  This sharp change arises from
specific qubit pairs with exceptionally low simultaneous RB process fidelity.
On \verb|ibm_fez| the pair of qubits \(\{71,72\}\) exhibits a process fidelity of
approximately \(0.063\) versus a mean of \(0.977\) across the chain, and this is
attributed to a two-qubit gate error exceeding \(2\times 10^{-1}\) and a readout
error of qubit~72 around \(0.456\), several orders of magnitude larger than the
device median. On \verb|ibm_kingston| the pair \(\{113,114\}\) has a process
fidelity near \(0.659\) compared with a chain mean of \(0.992\); here the CZ
error is above \(2\times 10^{-1}\) and the qubit~113 has a readout error around
\(0.329\). These weak links dominate the product in the layer fidelity and
therefore raise the EPLG sharply at the associated chain lengths.

Physically, the layer fidelity is a product of the process fidelities of each
two-qubit gate and idle qubit in the disjoint layer, so it decays approximately
exponentially with the number of entangling operations.  For small chains, all IBM
devices maintain fidelities above $80\%$, corresponding to EPLG values on the
order of $10^{-3}$–$10^{-2}$.  As the chains grow, cumulative crosstalk and
decoherence degrade performance, but the magnitude of the drop depends on the
uniformity of gate calibrations across the chip.  The abrupt jumps on
\verb|ibm_fez| and \verb|ibm_kingston| highlight that a single high‑error link
can dominate the fidelity of a long chain.  IBM Heron r3 device architectures
such as \verb|ibm_pittsburgh| offer more uniform gate quality, leading to smaller EPLG values throughout than others.  This emphasizes
the importance of calibrating all couplings uniformly and avoiding known bad
edges when mapping circuits. More EPLG results can be found at Table~\ref{tab:eplg-summary}. 
We note that for certain devices, the benchmark would fail to find a long qubit chain even if the device has a sufficient number of qubits, and this is because some of the qubit pairs are not connected (for example,  qubit pair 43-44 and 49-50 in \texttt{iqm\_emerald}).

We remark that, while trapped-ion processors often exhibit long coherence and high-fidelity entangling operations, our current EPLG results on \texttt{quantinuum\_h2\_2} should be interpreted with several protocol-specific caveats: (i) Because EPLG-10/20 are extracted from an $N=50$ run, the resulting values can be conservative on shuttling architectures where non-selected ions may still undergo transport and idle periods during circuit execution~\cite{pino2021demonstration,moses2023race};(ii) EPLG inserts barriers by default to enforce synchronized sublayer execution as in the canonical protocol~\cite{mckay2023benchmarking}. On transport-based devices this can induce additional idle time and hence larger memory-error contributions; (iii) our budget-limited sampling (\texttt{shots}=200, \texttt{num\_samples}=1) increases statistical uncertainty; (iv) The truncated RB depth schedule used here provides a less constrained decay fit, which can bias EPLG estimates when the observed decay is noisy~\cite{palacio2025parameter}, and we therefore regularize the RB fit by constraining the offset parameter $b$. To this end, we regard the present \texttt{quantinuum\_h2\_2} EPLG values as preliminary and expect more reliable estimates with deeper depth schedules and increased sampling in a future dataset release.

To compute an aggregated normalized EPLG score for inclusion in the overall
Metriq score (Section~\ref{sec:metriq-score}), we normalize EPLG values at
representative chain lengths against a baseline device $d^{\mathrm{base}}$ and
then average. For each chain length $\ell \in \{10, 20, 50, 100\}$ and device
$d$, the normalized component is
\begin{equation}
    S_{\mathrm{EPLG}(\ell)}(d) = 100 \frac{\mathrm{EPLG}(\ell, d^{\mathrm{base}})}{\mathrm{EPLG}(\ell, d)},
\end{equation}
where the ratio is inverted because lower EPLG values indicate better
performance. If a device does not support circuits of size $\ell$ or if no data
were collected at that size, we set $S_{\mathrm{EPLG}(\ell)}(d) = 0$. Let
$A(d)\subset \mathcal{L}$ be the subset of chain lengths for which device $d$
has nonzero normalized components $S_{\mathrm{EPLG}(\ell)}(d)>0$. To emphasize
larger problem scales, we assign nonnegative scale weights $w_\ell$ with
$\sum_{\ell\in\mathcal{L}} w_\ell = 1$; in this work we take $ w_\ell =
\frac{\ell}{\sum_{\ell'\in\mathcal{L}} \ell'}$ with $ (\ell\in\mathcal{L}) $. We
then define the aggregate EPLG Score as a weighted harmonic mean with a
weight-aware coverage penalty factor that reduces when measurements at larger
(higher-weight) chain lengths are missing:
\begin{equation}
    \label{eq:EPLGscore_definition}
    \textbf{EPLG~Score}(d)
    :=
    \sum_{\ell\in A(d)} w_\ell \cdot 
    \frac{\sum_{\ell\in A(d)} w_\ell}{
    \sum_{\ell \in A(d)} \frac{w_\ell}{S_{\mathrm{EPLG}(\ell)}(d)}
    }.
\end{equation}
For uniform weights $w_\ell = 1/|\mathcal{L}|$, this reduces to the unweighted
harmonic-mean definition with the $|A(d)|/|\mathcal{L}|$ coverage penalty. A
score of 100 corresponds to baseline performance; higher values indicate better
(lower-error) gate layers. While we adopt the weighting in
Eq.~\eqref{eq:EPLGscore_definition} to report EPLG Scores in this work, we
encourage readers to recompute the score using alternative weighting choices
that better match their preferred emphasis across chain lengths.

Table~\ref{tab:eplg-summary} reports the approximate scores along with each
processor’s maximum qubit count
and qubit pairs identified with significantly low process fidelity.

\begin{table}[H]
  \centering
  \begin{tabular}{lcccccccc}
    \toprule
    \textbf{Device} & \textbf{Qubits} & \makecell{EPLG-10\\ ($\times 10^{-3}$}) &\makecell{EPLG-20\\ ($\times 10^{-3}$}) &\makecell{EPLG-50\\ ($\times 10^{-3}$}) &  \makecell{EPLG-100\\ ($\times 10^{-3}$}) & \makecell{\textbf{EPLG score} \\ (normalized)} & \textbf{Low‑fidelity pair index} \\
    \hline
    \texttt{ibm\_boston} & 156 &1.64 &  2.16 &  2.45 &   3.08   & 338.40 &  None \\
    \texttt{ibm\_pittsburgh} & 156 & 2.43 &  2.58 &  3.10  & 4.13  & 258.53  & None \\
    \texttt{ibm\_marrakesh} & 156 & 2.97 &  3.80  & 4.14  & 5.06   & 201.11  & None \\
    \texttt{quantinuum\_h2\_2} & 56 & 1.80 & 1.76  & 1.73   &  --   & 181.53 &  None \\
    \texttt{ibm\_kingston} & 156 & 2.37 &  2.62  & 3.93 & 9.14 & 149.27 & $[113,114]$ \\
    \texttt{ibm\_torino} & 133 & 4.88 &   5.41 &   9.21  &  10.79 & 100.00  & None \\
    \texttt{ibm\_fez} & 156 & 5.29 &  5.98   &7.24 & 44.27   & 37.30  & $[71,72]$ \\
    \texttt{iqm\_emerald} & 54 & 4.53 & 8.10  &  --   &  -- &  12.75 & None \\
    \texttt{rigetti\_ankaa\_3} & 82 & 0.21 & 25.60  &   58.17  &  -- & 8.65  & None \\
    \texttt{iqm\_garnet} & 20 & 8.23 & 9.18  &  --   &  -- & 9.84  & None \\
    \texttt{wukong\_72} & 72 & $*$ & $*$ & $*$ &   -- & 0.00 & None \\
    \bottomrule
  \end{tabular}
  \caption{
  Average EPLG scores for selected devices. The EPLG score definition is in
  Eq.~\eqref{eq:EPLGscore_definition}. Note that values from the
  \texttt{wukong\_72} device were not able to be run
  due to device compilation issues, and are represented with the $*$ symbol. Entries for unsupported qubit number by the device are labeled by $-$ in tables of the this paper.  Data captured from
  November 2025 to February 2026.   \label{tab:eplg-summary}
  }
\end{table}

%------------------------------------------------------------------------------%
\subsubsection{Mirror circuits}
\label{sec:benchmark-mirror}
%------------------------------------------------------------------------------%
Mirror circuit benchmarking is a scalable and efficiently verifiable method to
evaluate the performance of quantum processors~\cite{proctor2021measuring, proctor2022scalable}. The central idea is to transform
a base quantum circuit $C$ into a family of mirror circuits $\{M_C\}$ that are
at least as hard to execute as $C$, but whose outputs are efficiently
predictable. Each mirror circuit is constructed by preparing qubits in random
Pauli eigenstates, running $C$ that has $d$ layers, applying a randomly chosen
Pauli layer $Q$, executing a quasi-inverse circuit $\tilde{C}^{-1}$ with the
same $d$ layers, and measuring in the initial basis. By design, every mirror
circuit has a unique correct output bit string, and the success probability $S$
of observing this bit string provides a direct measure of how well the processor
implements $C$. The polarization, defined as $P=(S-2^{-w})/(1-2^{-w})$ for width
$w$ (number of qubits), conveniently rescales $S$ to remove the uniform-random
baseline. Circuit mirroring is sensitive to realistic errors, including
crosstalk and coherent or biased stochastic noise, while remaining flexible to
either randomized or structured base circuits and scalable to large widths and
depths without exponential classical verification.

Our implementation follows Ref.~\cite{proctor2021measuring} by generating ensembles of
randomized mirror circuits that respect device connectivity and native gate sets. In
keeping with the quasi-inverse construction, we focus on Clifford layers
interleaved with Pauli layers, and we sample two-qubit interactions according to
a tunable per-layer probability while applying single-qubit Clifford gates to
idle qubits. This yields controllable two-qubit gate density and enables
systematic sweeps over circuit shape $(w,d)$ with minimal verification overhead.
Below we show an example configuration schema:
\begin{verbatim}
"benchmark_name": "Mirror Circuits",
"width": 10,
"num_layers": 16,
"shots": 1000,
"two_qubit_gate_prob": 0.5,
"num_circuits": 10
\end{verbatim}
The \texttt{width} parameter sets the number of qubits $w$; \texttt{num\_layers}
specifies the depth $d$ of the base half-circuit $C$ so that the mirror
benchmark depth is approximately $2d$ up to a constant overhead from state
preparation, the central Pauli layer, and measurement; \texttt{shots} is the
number of repetitions used to estimate $S$ (we typically use $10^3$, consistent
with prior work); \texttt{two\_qubit\_gate\_prob} controls the probability that a
given layer contains one or more two-qubit gates, thereby tuning the two-qubit
gate density that strongly influences performance; and \texttt{num\_circuits} sets
the number of independently sampled mirror circuits at the same shape, which
averages over circuit-to-circuit structural variability. The benchmark outputs
the success probability $S$, the polarization $P$ as well as the binary variable
that depends on whether $S$ is above the threshold $1/e$. Relative to the
original study, our current runs emphasize randomized mirror circuits rather
than periodic ones, and we automate sampling, execution, and polarization
aggregation in a common pipeline. 

\begin{figure}
    \centering
    \includegraphics[width=0.8\linewidth]{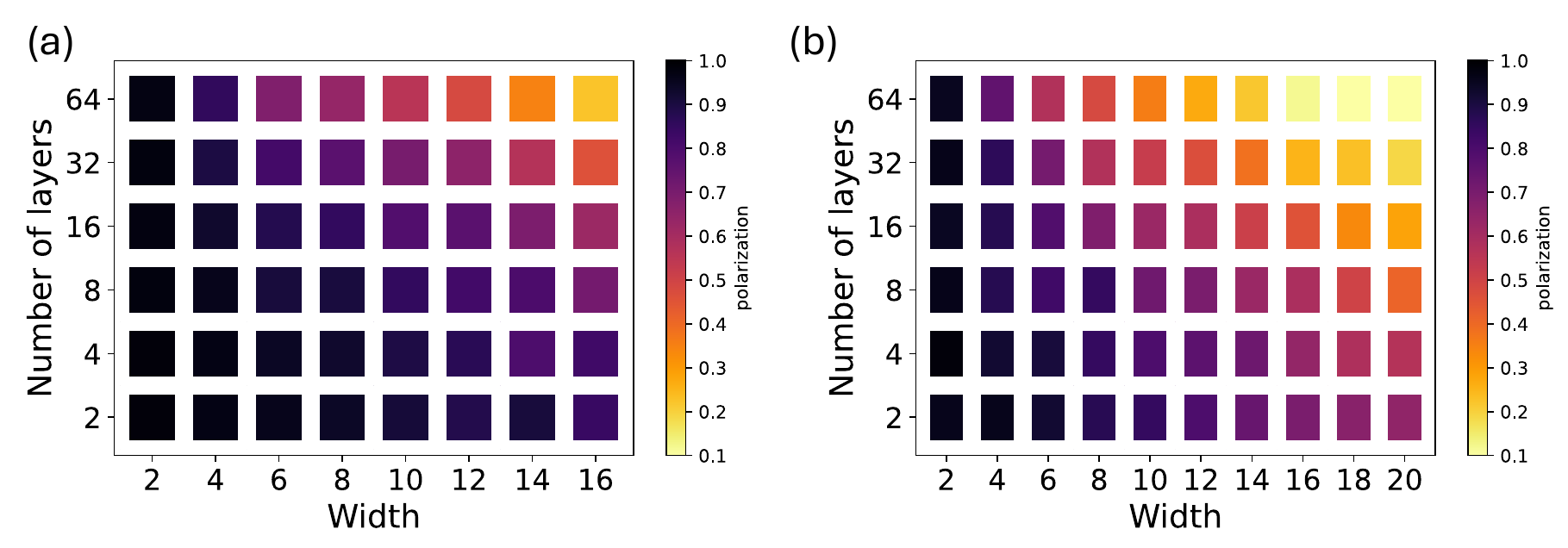}
    \caption{Mirror circuit benchmarking on \texttt{ibm\_fez}. Panel (a) shows a
    noisy simulation that uses realistic device-calibrated gate and readout
    error parameters, and panel (b) shows experimental data from the same
    backend. Each square represents a circuit shape with width (horizontal axis)
    and number of layers (vertical axis); color encodes the polarization
    $P=(S-2^{-w})/(1-2^{-w})$. The color bar is identical for both panels,
    enabling direct comparison between simulated and measured performance. Data were taken in December 2025.}
    \label{fig:mirror-fez}
\end{figure}

We next present the hardware results of the mirror circuit benchmarking.
Fig.~\ref{fig:mirror-fez} illustrates mirror circuit results on
\texttt{ibm\_fez}, plotting polarization versus width and the number of layers.
The polarization decays with increasing depth and width as expected from error
accumulation, yet remains appreciable at moderate shapes, indicating the device
can reliably execute circuits of those sizes. These measurements offer an
at-a-glance capability profile and exemplify how the mirror protocol converts
arbitrary workloads into efficiently verifiable tasks while preserving
sensitivity to error structure.

We next define a scalar score that summarizes the performance of a quantum
processor using mirror circuits. The benchmark uses a fixed panel of width-depth
pairs
\begin{equation}
    \mathcal P = \{(8, 64), (16, 32), (24, 16), (32, 8), (64, 4), (128, 2)\},
\end{equation}
so the panel size is $M=6$. For each $(w_i,d_i)\in \mathcal P$, we generate $\texttt{num\_circuits}=10$
randomized mirror circuits with a fixed two-qubit gate probability $p_{2q}=0.5$.
Each circuit is executed with a fixed number of shots $N_{\text{shots}}$ ($N_{\text{shots}}=1000$ for results in this work). The
expected output bitstring is computed by Clifford conjugation of the middle
Pauli layer and therefore does not require a classical state-vector simulation,
even for large $w_i$. For a given panel point, let $s_i$ be the total number of
shots (pooled over the $\texttt{num\_circuits}$ circuits) that match the expected bitstring, and let
$c_i$ be the total number of shots collected. The empirical success probability
is simply $\hat{p}_i = \frac{s_i}{c_i}.$ A device that outputs uniformly random
bitstrings would succeed with probability $2^{-w_i}$, so we define the
baseline-corrected polarization
\begin{equation}
    \pi_i = 
    \max\left\{0, \frac{\hat{p}_i - 2^{-w_i}}{1 - 2^{-w_i}}\right\} = 
    \max\left\{0,\frac{2^{w_i} \hat{p}_i - 1}{2^{w_i} - 1}\right\},
\end{equation}
which always satisfies $0 \le \pi_i \le 1$. If a device does not have at least
$w_i$ usable qubits, we set $\pi_i = 0$ for that panel point; this makes the
score sensitive to both performance and scale. Note that one can use the full
Hamming-distance distribution of output bitstrings to compute higher-order
contributions and form an effective polarization, as shown in
Ref.~\cite{proctor2022scalable}. That approach exploits information beyond
exact-bitstring matches and enables depth-dependent fits to extract a layer
error rate, while the present score uses the $k=0$ term only, which simplifies
interpretation and keeps the metric uniform across widths without model fitting.

The mirror circuit benchmark evaluates a fixed panel of $M=6$ instances at
circuit widths $n_i\in\{8,16,24,32,64,128\}$. For each panel point $i$ we compute
a polarization $\pi_i\in[0,1]$ (defined above).

To obtain a single benchmark-level raw value, we aggregate across widths using
weights proportional to circuit width (Eq.~\eqref{eq:width-weights}):
\begin{equation}
    \textbf{MC Score}
    := \sum_{i=1}^{M} \alpha_i\, \pi_i,
    \qquad
    \alpha_i := \frac{n_i}{\sum_{j=1}^{M} n_j}.
    \label{eq:mc-width-aggregate}
\end{equation}
For the fixed panel $\{8,16,24,32,64,128\}$, this corresponds to weights
$(1,2,3,4,8,16)/34$, thereby emphasizing performance at larger widths. Since
$\sum_i \alpha_i=1$ and $\pi_i\in[0,1]$, we have $\text{MC Score} \in[0,1]$.
Devices that can run all panel points with high fidelity will achieve values
close to $1$. Devices that can run only smaller-width points, or that approach
the random baseline at large width, obtain smaller values because the
corresponding $\pi_i$ are near zero (and, when a panel point cannot be executed,
we set its contribution to zero).

A conservative overall uncertainty can be reported as the standard error of the
(weighted) mean across the panel points. Table~\ref{tab:mc-scores} reports the
raw values $\bar v_{\mathrm{MC}}$; Table~\ref{tab:metriq-summary} reports the
corresponding normalized subscore $\mathrm{BS}_{\mathrm{MC}}$.

\begin{table}[H]
    \centering
    \begin{tabular}{lccccccccc}
        \toprule
        \textbf{Device} & \textbf{Qubits} & (8,64)& (16,32)& (24,16) & (32,8) & (64,4) & (128,2) & \textbf{Verbatim mode} & \textbf{MC Score} \\
        \midrule
        \texttt{ibm\_boston} & 156 & 0.7477 & 0.4952 & 0.4317 & 0.4707& 0.2661
        &0.1122  & No & 0.260000  \\
        \texttt{quantinuum\_h2\_2} & 56 &  0.7227 & 0.6898  & 0.7470  & 0.8213 &
        -- & -- & No & 0.224368  \\
        \texttt{ibm\_kingston} & 156 & 0.5194 & 0.3871& 0.3337 & 0.3309& 0.1528
        & 0.0257 & No & 0.154468 \\
        \texttt{ibm\_pittsburgh} & 156 & 0.6042 & 0.4613 & 0.3309 & 0.2250 &
        0.0164 & 0.0145 & No & 0.111256 \\
        \texttt{ibm\_marrakesh} & 156 & 0.4949 & 0.4846 & 0.4014& 0.1285& 0.0371
        & 0.0027 & No & 0.103597 \\
        \texttt{ibm\_fez} & 156 & 0.4507 & 0.2624 & 0.2868& 0.0944 & 0.0266 &
        0.0012 & No & 0.071926 \\
        \texttt{ibm\_torino} & 133 & 0.3172 & 0.2757 & 0.1268 & 0.0336 & 0.0037
        & 0.0 & No & 0.041559 \\
        \texttt{iqm\_emerald} & 54 & 0.0210 & 0.0038 & 0.0310 &
        0.0259 & -- & -- & Yes & 0.006624 \\
        \texttt{iqm\_garnet} & 20 & 0.0337 & 0.0043 & -- & -- & --
        & -- & Yes & 0.001244 \\
        \texttt{rigetti\_ankaa\_3} & 82  & 0.0040 & 0.0 & 0.0 & 0.0 &
        0.0 & -- & Yes & 0.000118 \\
        \texttt{wukong\_102} & 102 & 0.0039 & 0.0 & 0.0 & 0.0 & 0.0 & -- & No &
        0.000115 \\
        \texttt{wukong\_72} & 72 & 0.0006 & 0.0 & 0.0 & 0.0 & 0.0 & -- & No & 0.000018
        \\
        \bottomrule
    \end{tabular}
    \caption{Mirror circuits polarization values for various parameter
    sets $(w,d)$ where $w$ corresponds to \texttt{width} and $d$ corresponds to \texttt{num\_layers}, as well as score results for various quantum
    processors. All data were taken in the time window of December 2025 to February 2026.} 
    \label{tab:mc-scores}
\end{table}

%------------------------------------------------------------------------------%
\subsubsection{Circuit Layer Operations Per Second (CLOPS)}
\label{sec:benchmark-clops}
%------------------------------------------------------------------------------%
As a complement to the more quality-focused system-level benchmarks previously discussed,
the Circuit Layer Operations Per Second (CLOPS) benchmark measures the sustained rate at which a device can execute layers
of gates within a randomized EPLG-like circuit. This provides insight into its computational throughput and real-time performance. Originally introduced by IBM~\cite{wack2021quality}, CLOPS has also been adopted by other vendors \cite{vanderschoot2023clops, abdurakhimov2024clops} and, to our knowledge, is the only speed-oriented benchmark proposed for current and near-term quantum hardware. Crucially, CLOPS is not only influenced by low-level gate times but also captures higher-level factors such as
circuit compilation efficiency, queueing delays, and classical-quantum communication overheads.
This helps users understand the practical speed of quantum computers on industrially relevant workloads.

Our implementation builds on the updated protocol~\cite{wack2026clops,qiskit2025benchmarking}, which aligns CLOPS with the hardware-aware layering defined in the EPLG benchmark (Section~\ref{sec:benchmark-eplg}). The benchmark defines a circuit template over $N$ qubits and $L$ layers, where each layer contains a selection of two-qubit gates on disjoint pairs of qubits, followed by parameterized single-qubit rotations. We instantiate $M$ circuits of this form, assigning random angles to the parameterized rotations, and measure $S$ shots per instantiated circuit. This structure represents typical near-term workloads such as variational algorithms~\cite{cerezo2021variational} that involve repeated layers of parameterized operations. Additionally, the randomization mirrors techniques used in error mitigation, such as Pauli twirling~\cite{wallman2016twirling}, which require repeated execution of randomized circuit instances. As with EPLG, the choice of two-qubit pairs is more typical of devices with nearest-neighbor connectivity, and thus might not measure typical performance for all architectures.

Given this workload of circuits to run, the CLOPS benchmark score is defined as the total number of layers executed divided by the total execution time $T_{tot}$
\begin{equation}
  \mathrm{CLOPS} = \frac{ L M S} {T_{tot}}.
\end{equation}
Here, $T_{tot}$ is the execution time provided by the cloud runtime job result data. It measures from the time the first circuit in the workload starts running to the time the last circuit run completes. This is meant to ignore queuing times, but still capture the end-to-end performance of the entire cloud stack running the circuit. However, in practice, the exact definition of $T_{tot}$ can be ambiguous and may differ across platforms. Indeed, many platforms do not report any timing data at all, and for those that do, the reported timing may include or exclude different components of the execution time, such as compilation time, recalibration time, waiting times, etc. This makes it difficult to compare CLOPS scores across platforms in a truly platform-agnostic manner, but we hope this benchmark motivates providers to report more detailed timing data in the future, and to clarify what is included in the reported execution time. This timing definition can also be sensitive to any initial startup work that is otherwise amortized over the runtime of the entire workload. Our implementation thus also reports a steady-state CLOPS score, which excludes the timing and work done by the first job for platforms which report granular timing data. As of this writing, only Quantinuum NEXUS platform and IBM Quantum Cloud report execution timing data, and only IBM provides the granular sub-job execution data. 

In addition to the subtleties in timing, there are also platform-specific capabilities that can impact the CLOPS score and thus real-world performance. We expose some as configuration options in the benchmark, even if they are only supported by a single vendor. The description of the implementation above corresponds to the baseline \texttt{instantiated} mode of the benchmark. In this mode, all circuits are fully instantiated locally and all are dispatched to the cloud in a single aggregate job to execute. As a result, the maximum workload size is limited by the cloud provider's API limits. We also define a \texttt{parameterized} mode, which sends the circuit template with parameters as placeholders alongside the list of parameter values for each instantiation. The cloud runtimes can leverage this information to do one-off compilation, lowering and device preparation knowing the circuit structure, and then just do parameter instantiation per-circuit, which can be much faster. Similarly, a \texttt{twirled} mode eliminates the parameters entirely, and uses native twirling capabilities to effectively instantiate the random single-qubit gates at execution time using twirling primitives. Both of these advanced modes can support larger workloads to better showcase steady-state performance, but are currently only supported on IBM Quantum Cloud. We hope to extend support to other platforms in future versions of \mgym.

As of this publication, we only have results for IBM cloud devices, as listed in Table~\ref{tab:clops-scores-transposed} using the \texttt{twirled} mode. An example configuration schema for these runs is:
\begin{verbatim}
"benchmark_name": "CLOPS",
"num_qubits": 100,
"num_layers": 100,
"num_circuits": 1000,
"shots": 100,
"mode": "twirled",
"two_qubit_gate": "cz",
\end{verbatim}
where \texttt{mode} is one of instantiated, parameterized, and twirled, and \texttt{two\_qubit\_gate} is the two-qubit gate to use in the start of each layer. 

\begin{table}[H]
    \centering
    \begin{tabular}{lcc}
        \toprule
        \textbf{Device} &  \textbf{CLOPS} \\
        \midrule
        \texttt{ibm\_boston}     & 360{,}573 \\
        \texttt{ibm\_fez}        & 368{,}905 \\
        \texttt{ibm\_kingston}   & 362{,}371 \\
        \texttt{ibm\_marrakesh}  & 353{,}641 \\
        \texttt{ibm\_pittsburgh} & 358{,}612 \\
        \texttt{ibm\_torino}     & 345{,}669 \\
        \bottomrule
    \end{tabular}
    \caption{CLOPS scores across tested IBM devices. All runs used $N=100$ qubits, $L=100$ layers, $M=1000$ circuits, and $S=100$ shots.
    CLOPS being evaluated at a single width ($n=100$) makes its width aggregation trivial.}
    \label{tab:clops-scores-transposed}
\end{table} 
Preliminary results were obtained on \texttt{quantinuum\_h2\_2}, but the API provided timing data included downtime and queuing due to job sharing across the device. These results, therefore, measure the cloud platform more than the device performance, and new results need to be collected as more granular timing data is available. For the IBM devices with reliable timing data, the raw CLOPS scores are normalized against the baseline \texttt{ibm\_torino} device to compute their contribution to the aggregate Metriq score. For other hardware vendors to be fully supported by this benchmark, their cloud APIs would need to be updated to expose isolated, granular execution timing that strictly excludes classical queueing, job sharing, and platform downtime. Additionally, because CLOPS measures the execution of parallel layered operations, the resulting score is, to first order, independent of the total number of qubits, acting instead as a proxy for device parallelization. However,  in order to rigorously compare throughput across heterogeneous devices as they add timing support at varying qubit counts, future work must characterize how CLOPS scales across different processor sizes to identify potential control-system tradeoffs or bottlenecks.

While integrating CLOPS into \mgym represents a significant step towards a cross-platform speed benchmark, we recognize that platform-specific timing definitions and feature sets currently limit comparability. Furthermore, due to the orders-of-magnitude differences in native gate speeds across hardware modalities (e.g., superconducting circuits vs. trapped ions), raw CLOPS scores will naturally diverge. Therefore, we posit that the primary value of this benchmark lies in tracking the evolution of a specific platform's performance over time, which captures improvements in compiler efficiency, control electronics, and runtime architecture, rather than serving as a simplistic ranking between disparate hardware architectures. We aim to extend support for advanced compilation modes to additional providers as their APIs mature, enabling more equitable comparisons of best-effort performance. Finally, while CLOPS targets the NISQ and early fault-tolerant regimes, we anticipate that as hardware capabilities advance, the focus will shift towards logical-level throughput. In this context, we look forward to incorporating metrics such as the recently proposed Quantum Logical Operations Per Second (QLOPS)~\cite{kong2025qlops} to benchmark the speed of error-corrected computation.
%------------------------------------------------------------------------------%
\subsection{Application-inspired circuit benchmarks}
\label{sec:benchmark-application}
%------------------------------------------------------------------------------%

In addition to system-level metrics, the \mgym suite includes a set of
application-inspired circuit benchmarks derived from representative
quantum algorithmic workloads. Some of these benchmarks build on
previously established protocols, while others formalize algorithmic
primitives into benchmarking protocols for the first time within a
standardized, cross-platform framework.

These benchmarks assess structured quantum circuit execution by capturing the topology, depth, and error sensitivity of representative quantum algorithm classes under controlled and reproducible conditions. They do not measure end-to-end application performance, since they exclude classical preprocessing and post-processing, hybrid orchestration, error mitigation, and time-to-solution analysis. Instead, they isolate the quantum processing stage. Within Metriq, their purpose is to compare hardware systematically across representative workload classes, while recognizing that the results remain circuit-level and task-dependent rather than absolute application-level measures.

%------------------------------------------------------------------------------%
\subsubsection{Quantum machine learning (QML) kernel}
\label{sec:benchmark-qml}
%------------------------------------------------------------------------------%
As part of our application‑level benchmarks, \mgym includes a Quantum Machine
Learning (QML) kernel task that evaluates a processor’s ability to compute
kernel matrix elements for machine‑learning applications~\cite{thanasilp2024exponential,
schuld2021supervised}. Quantum kernel methods embed classical data points
$x\in\mathbb{R}^d$ into a high–dimensional Hilbert space via a feature map
$U(x)$ acting on an $n$‑qubit register.  Given two data points $x_i$ and~$x_j$,
the resulting states $\ket{\Phi(x_i)}=U(x_i)\ket{0}^{\otimes n}$ and
$\ket{\Phi(x_j)}=U(x_j)\ket{0}^{\otimes n}$ define a kernel
\begin{equation}
 k(x_i,x_j)=\bigl|\langle \Phi(x_i)\,|\,\Phi(x_j)\rangle\bigr|^2,
 \label{eq:qkernel}
\end{equation}
which constitutes a single entry of the Gram matrix used in a classical kernel method such as support vector machines.  A quantum device can estimate Eq.~\eqref{eq:qkernel} by preparing $\ket{\Phi(x_i)}$ and $\ket{\Phi(x_j)}$ and then performing a swap test or, alternatively, by building a combined circuit whose second half implements $U(x_j)^\dagger$ on the same register.  When $x_i=x_j$, the circuit ideally returns the ground state $\ket{0}^{\otimes n}$ and Eq.~\eqref{eq:qkernel} equals one.

The QML kernel benchmark implemented in \mgym adopts a parametrized energy‑efficient feature map inspired by Ref.~\cite{yamauchi2024parameterized}.  For $n$ qubits, the feature map circuit $U_{\mathrm{ZZ}}(x)$ begins with a layer of Hadamard gates on each qubit and a layer of single–qubit $R_z$ rotations parameterized by the classical feature vector $x$:
\begin{equation}
     U_{\mathrm{ZZ}}(x) = \mathcal{E}(x)\prod_{k=0}^{n-1}R_z(x_k)_k \prod_{k=0}^{n-1}H_k,
     \label{eq:zz-feature}
\end{equation}
where $\mathcal{E}(x)$ denotes a sequence of two entangling layers.  Each layer applies controlled‑NOT (CNOT) gates between nearest‑neighbour pairs followed by a conditional $R_z$ rotation on the target qubit; the rotation angle depends on the product of rotated angles $(\pi-x_{i})(\pi-x_{j})$ for the pair $\{i,j\}$.  In code, the first entangling layer acts on qubit pairs $(0,1),(2,3),\ldots$, while the second layer acts on $(1,2),(3,4),\ldots$.  This pattern mirrors the energy‑efficient quantum kernel proposed in Ref.~\cite{yamauchi2024parameterized} and yields a relatively shallow circuit with long‑range correlations.

To estimate the kernel $k(x,x)$, the benchmark constructs an inner‑product
circuit that combines two copies of $U_{\mathrm{ZZ}}(x)$, with the second copy
applied in reverse order.  In Qiskit, this is accomplished using the
\verb|unitary_overlap| generator, which returns a circuit that prepares
$U(x)^\dagger U(x)$ on a single register.  Explicitly,
\begin{equation}
\ket{\psi}\;=\;U_{\mathrm{ZZ}}(x)^{\dagger}\,U_{\mathrm{ZZ}}(x)\ket{0}^{\otimes n}
\end{equation}
so that, in the absence of noise and decoherence, one expects $\ket{\psi}=\ket{0}^{\otimes n}$ and therefore $k(x,x)=1$.  In the benchmark, we randomly sample a parameter vector $x\in[0,2\pi]^n$ and assign it to both halves of the circuit so that the ideal overlap is unity.  With measurements on every qubit, the probability of measuring the all‑zero bit‑string defines the accuracy of the circuit and provides an estimate of the kernel value. The benchmark is configurable through the number of qubits $n$ and the number of shots used. An example configuration for a 50‑qubit instance is:
\begin{verbatim}
"benchmark_name": "QML Kernel",
"num_qubits": 50,
"shots": 1000
\end{verbatim}

\begin{figure}
    \centering
    \includegraphics[width=0.8\linewidth]{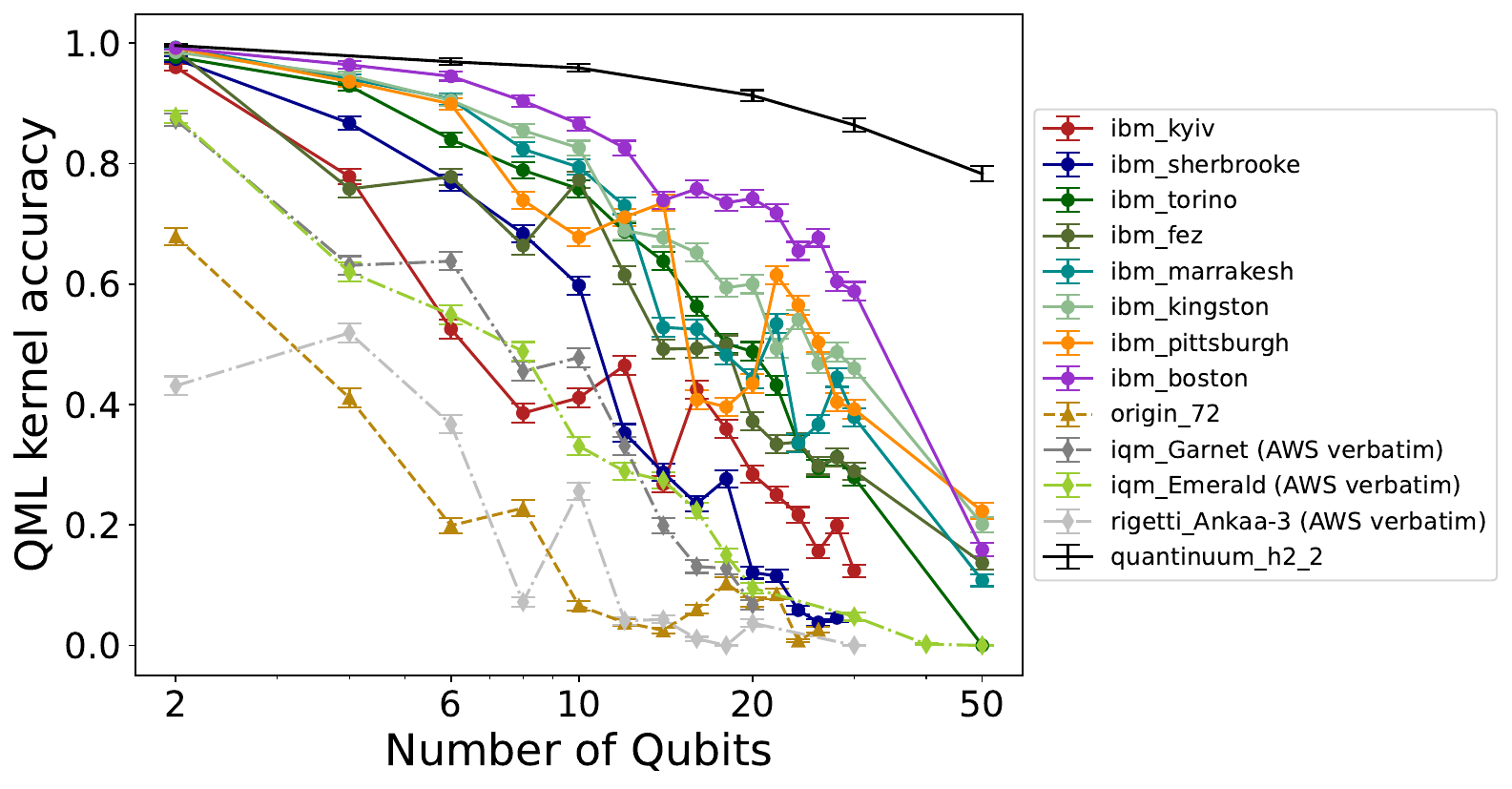}
    \caption{
    QML kernel accuracy for a range of qubit counts on different hardware
    platforms. The number of shots is 1000. The figure distinguishes between
    circuits compiled normally and those executed with verbatim compilation on
    AWS Braket due to missing barrier support at the time of data collection (see
    Section~\ref{sec:benchmarking-compilation} for details).}
    \label{fig:QMLKernel_results}
\end{figure}

The QML kernel benchmark occupies a unique place in our benchmarking suite
because it directly probes the hardware's performance on a machine‑learning task
and realistic quantum‑enhanced data processing.  Unlike low‑level metrics such
as single‑gate fidelity or relaxation times, quantum kernel evaluation requires
coherent implementation of multi-qubit circuits with structured entanglement and
parameterized rotations. Our implementation uses an energy‑efficient feature map
to reduce circuit depth while still generating non‑trivial entanglement, thereby
making the task accessible to contemporary hardware.  The simple accuracy metric
enables straightforward comparison between platforms and provides a gateway to
more sophisticated tasks, such as full kernel matrix estimation or training
support vector machines.

A notable characteristic of the QML kernel benchmark is its compatibility with sampling-based quantum error mitigation techniques.
The QML kernel evaluates the probability of the all-zero outcome, $p(0^n)=\langle \psi|\Pi_{0^n}|\psi\rangle$, i.e., the expectation
value of the projector $\Pi_{0^n} = |0\rangle\!\langle 0|^{\otimes n}$. While $\Pi_{0^n}$ can be written as a sum of Pauli-$Z$ strings,
this expansion contains $2^n$ terms, so Pauli-term expectation-estimation strategies are inefficient at scale. Consequently, methods
that operate directly on sampled outcome distributions (e.g., readout error mitigation) are the most natural fit for this benchmark.
This makes the QML kernel benchmark a valuable test case for sampling-based mitigation on near-term quantum devices.

We then show the hardware results for QML kernel benchmarking.
Fig.~\ref{fig:QMLKernel_results} shows the measured QML kernel accuracy as a
function of the number of qubits for several superconducting and trapped‑ion
devices.  Each point corresponds to the benchmark described in
Section~\ref{sec:benchmark-qml}: a random parameter vector is used to build an
$n$‑qubit energy‑efficient feature map; two copies of this circuit are composed
to estimate the squared inner product, and the probability of measuring the
all‑zero bit‑string is reported as the QML kernel accuracy. Error bars denote
binomial standard errors.  The general trend across platforms is a monotonic
decline in accuracy with increasing qubit count: most devices achieve near‑ideal
performance ($\gtrsim 0.95$) for two qubits and drop below $0.6$ by about twenty
qubits. Beyond roughly fifty qubits the kernel accuracy of all platforms
approaches zero, which justifies our choice to truncate the benchmark at fifty
qubits.
Devices executed via AWS Braket (including \texttt{iqm\_garnet}, \texttt{iqm\_emerald}, and \texttt{rigetti\_ankaa\_3}) were
restricted to verbatim compilation because barrier operations were not supported
at the time of data collection; as a result, additional compiler‑inserted gates
may have degraded performance.  We refer to
Section~\ref{sec:benchmarking-compilation} for a detailed discussion of compilation
settings and their impact.

To provide a single figure of merit, we evaluate QML Kernel at widths
$n\in\{10,20,30,50\}$. Let QMLK-n denote the accuracy at $n$ qubits. We
aggregate across widths as
\begin{equation}
  \textbf{QMLK Score}
  := \sum_{n\in\{10,20,30,50\}} \frac{n}{110}\, \text{QMLK-n},
  \label{eq:qmlk-width-aggregate}
\end{equation}
(i.e., weights $(1,2,3,5)/11$).
If a device does not support circuits of size~$n$, 
or if no data were collected at that size due to device failure or device retirement, 
we assign the corresponding term a value of zero.  This definition focuses on the scaling behavior at intermediate circuit sizes and avoids over‑weighting trivial several‑qubit instances.  Table~\ref{tab:qmlk-scores} summarizes the approximate scores extracted from Fig.~\ref{fig:QMLKernel_results}.  The ``verbatim'' column indicates whether a device was executed on AWS Braket with verbatim compilation.

The QMLK Score highlights pronounced differences across platforms. 
IBM Heron architectures such as \texttt{ibm\_kingston} consistently show better performance than Eagle
architectures such as \texttt{ibm\_sherbrooke} due to their reduced two-qubit
gate error rates. The performance of device accessed through AWS Braket is likely
limited by compilation. %and control during data acquisition. 
These results demonstrate that current noisy devices can compute kernel entries with
reasonable accuracy for small data sets, but struggle as circuit width grows.

The significance of the QML kernel benchmark lies in its ability to bridge
theoretical quantum capabilities with practical applications.  By directly
measuring how faithfully a quantum processor can compute Gram matrix elements
for a machine‑learning task, the benchmark provides invaluable insights into how
effectively current devices can be leveraged for realistic quantum‑enhanced
learning.  This, in turn, informs both hardware development and algorithm design
for near‑term applications~\cite{yamauchi2024parameterized}.  The poor scaling observed in
Fig.~\ref{fig:QMLKernel_results} suggests that error‑mitigation techniques and
more compact kernel circuits will be essential to achieve practical quantum
utilities.

\begin{table}
    \centering
    \begin{tabular}{lccccccc}
        \toprule
        \textbf{Device} & \textbf{Qubits} & QMLK-10 &QMLK-20 & QMLK-30 & QMLK-50 & \textbf{Verbatim mode} & \textbf{QMLK Score} \\
        \midrule
        \texttt{quantinuum\_h2\_2} & 56 & 0.959 & 0.913 & 0.864 &0.783 & No & 0.844727 \\ 
        \texttt{ibm\_boston} & 156 & 0.866 & 0.742 & 0.588 & 0.159 & No & 0.446273 \\
        \texttt{ibm\_kingston} & 156  & 0.826 & 0.600 & 0.460 & 0.201 & No & 0.401000 \\
        \texttt{ibm\_pittsburgh} & 156 & 0.678 & 0.435 & 0.392 & 0.223  & No & 0.349000 \\
        \texttt{ibm\_marrakesh} & 156 & 0.794 & 0.443 & 0.379 & 0.108  & No & 0.305182 \\
        \texttt{ibm\_fez} & 156  & 0.773 & 0.372 & 0.289& 0.137 & No & 0.279000 \\
        \texttt{ibm\_torino} & 133 & 0.758 & 0.488 & 0.279 & 0 & No & 0.233727 \\
        \texttt{iqm\_emerald} & 54  & 0.331 & 0.095  & 0.048 & 0 & Yes & 0.060455 \\
        \texttt{iqm\_garnet} & 20  & 0.478& 0.068 & -- & -- & Yes & 0.055818	 \\
        \texttt{rigetti\_ankaa\_3} & 82 & 0.256 &0.037 & 0 & 0  & Yes & 0.030000 \\
        \texttt{wukong\_72} & 72  & 0.066 & 0.072 & 0.026 & 0 & No & 0.026182 \\
        \texttt{wukong\_102} & 102  & 0.014 & 0 & 0 & 0 & No & 0.001273 \\
        \midrule
        \multicolumn{4}{l}{\textit{Historical measurements:}} \\
        \texttt{ibm\_torino} (March 2025) & 133  &0.759 & 0.485 & 0.265 & $*$ & No & 0.229455 \\
        \texttt{ibm\_kyiv} & 127  & 0.411 &0.284 &0.124 & $*$& No & 0.122818 \\
        \texttt{ibm\_sherbrooke} & 127  & 0.598 & 0.121 & $*$ & $*$ & No & 0.076364 \\
        \bottomrule
    \end{tabular}
    \caption{QMLK Score for various quantum processors. Data
    on \texttt{iqm\_garnet}, and Rigetti devices were taken on June 2025 and \texttt{iqm\_emerald} on December 2025 to January 2026;
    \texttt{ibm\_sherbrooke} and \texttt{ibm\_kyiv} results were taken on March
    2025, while data on other IBM devices on November 2025 unless specified. The label $*$ means that data were not taken before the retirement of the online platform.}
    \label{tab:qmlk-scores}
\end{table}

%------------------------------------------------------------------------------%
\subsubsection{Wormhole-inspired teleportation (WIT)}
\label{sec:benchmark-wit}
%------------------------------------------------------------------------------%
The WIT benchmark is an application-inspired protocol inspired by holographic
duality and wormhole teleportation in the AdS/CFT correspondence. It is based on
the quantum circuit design from~\cite{shapoval2023towards}, which explores the
possibility of simulating gravitational physics, specifically traversable
wormhole dynamics, on quantum processors. In the AdS/CFT framework,
entanglement between two quantum systems is conjectured to be dual to a
geometric wormhole connecting two black holes. The WIT circuit implements a toy
model of this phenomenon by creating a highly entangled state, evolving it under
a carefully constructed Hamiltonian, and measuring the fidelity with which
quantum information is transmitted through the simulated wormhole geometry.

Our fixed implementation of the circuit operates on either 6 or 7 qubits. While
the original work by Shapoval \textit{et al.} allows for larger system sizes, we
restrict our implementation to these smaller scales for practical reasons. The
WIT protocol is susceptible to noise due to the long idle times experienced by
qubits during the multi-stage evolution, as many qubits remain idle while others
undergo operations, leading to significant decoherence accumulation. As a
result, near-term WIT performance is often dominated by idle-induced
decoherence, so the benchmark could be interpreted as a stress test of coherence
and scheduling rather than a balanced application metric at this stage. Current
quantum hardware struggles to achieve expectation values within even 10\% of the
ideal target without extensive error mitigation, and published results achieving
such fidelity typically employ sophisticated mitigation techniques. Scaling
beyond 7 qubits under present noise levels would yield expectation values too
close to the random baseline to provide a meaningful benchmarking signal. As
hardware fidelity improves, particularly with the advent of error-corrected
logical qubits, extending the WIT benchmark to larger system sizes will
typically become both feasible and valuable for probing coherence at scale.

The WIT circuit consists of several stages. First, three Bell pairs are prepared
using Hadamard gates followed by CNOT operations, creating maximal entanglement
between pairs of qubits. These entangled pairs represent the initial quantum
state before wormhole traversal. Next, the circuit simulates evolution under a
holographic Hamiltonian through a sequence of single-qubit rotation gates
($R_X$, $R_Z$) and two-qubit $R_{ZZ}$ entangling gates, with specific rotation
angles chosen to model the interaction coupling constant $g = \pi/2$. This
evolution is applied in three repeated layers, implementing the forward time
evolution.
At the midpoint of the circuit, the 6-qubit version applies a reset operation to
qubit 0, while the 7-qubit version uses a SWAP gate between qubits 0 and 6,
simulating the information insertion step in the wormhole teleportation
protocol. Following this, the circuit applies the time-reversed evolution: the
signs of the rotation angles are flipped, and three more layers of gates are
applied with the same structure but opposite rotations, effectively implementing
backwards time evolution. Finally, two critical parameterized $R_{ZZ}(\pi/2)$
gates couple qubits which represent the wormhole traversal interaction. The
circuit concludes with a measurement in the Pauli-$Z$ basis.

The benchmark reports a single metric: the expectation value of the measurement
outcome. In an ideal, noise-free scenario, the quantum state at the circuit's
end should be a well-defined computational basis state, yielding an expectation
value of 1.0. Deviations from this ideal value quantify the cumulative effects
of gate errors, decoherence, crosstalk, and readout imperfections throughout the
multi-stage protocol. The WIT benchmark configuration is straightforward:
\begin{verbatim}
"benchmark_name": "WIT",
"num_qubits": 7,
"shots": 8192
\end{verbatim}
The \texttt{num\_qubits} parameter must be either $6$ or $7$, corresponding to the two
circuit variants described above. The \texttt{shots} parameter specifies the
number of measurement repetitions used to estimate the expectation value; we
typically use several thousand shots (e.g., 8192) to obtain high statistical
precision for this single-qubit measurement.

The significance of the WIT benchmark lies in its role as a bridge between
fundamental theoretical physics and practical quantum computing applications.
Unlike traditional benchmarks that focus primarily on gate fidelities, coherence
times, or simple entanglement tests, WIT evaluates a device's ability to execute
complex, application-motivated quantum circuits. The protocol demands
simultaneous preparation of multiple Bell pairs across non-adjacent qubit pairs,
precise implementation of parameterized rotation gates with specific angles, and
coordinated multi-qubit entangling operations ($R_{ZZ}$ gates) that must
maintain phase coherence throughout. With approximately 70-80 total gates
organized in a multi-stage structure, the circuit requires maintaining quantum
coherence through moderately deep computation, followed by accurate single-qubit
readout. These stringent requirements make WIT a demanding end-to-end test of
device capabilities for quantum simulation applications. As quantum
simulation, particularly of condensed matter and high-energy physics
systems, emerges as one of the most promising near-term applications for
quantum computers, benchmarks like WIT that assess performance on physically
motivated circuits become increasingly important. Success on the WIT benchmark
indicates that a device is capable of executing the types of structured,
multi-stage quantum protocols needed for exploring quantum gravity models,
holographic correspondences, and other fundamental physics simulations.

Our implementation follows the circuit specifications
in Ref.~\cite{shapoval2023towards} with minor adaptations for cross-platform
compatibility. The primary difference from the reference work is our use of a
standardized circuit construction interface that automatically maps the logical
circuit to the device's native gate set and connectivity, ensuring consistent
behavior across multiple hardware providers. Additionally, we report uncertainty
estimates derived from shot noise statistics, enabling a more rigorous comparison
of results across different devices.

To contextualize WIT performance against circuit complexity, we compare the
measured expectation values to theoretical predictions based on simple
depolarizing noise models. For a circuit with an average two-qubit gate
fidelity $F_{\text{2q}}$, the expected final state fidelity decays
approximately as $F_{\text{circuit}} \approx
F_{\text{2q}}^{n_{\text{2q}}}$, where $n_{\text{2q}}$ is the number of two-qubit
gates. The 7-qubit WIT circuit contains 57 single-qubit gates and 24 two-qubit
gates (CNOT operations and $R_{ZZ}$ entangling gates). Assuming single-qubit
gate errors are negligible, we infer effective two-qubit gate fidelities as
$F_{\text{2q}} = E^{1/24}$, where $E$ is the measured expectation value.
Table~\ref{tab:wit-performance} lists the measured expectation values (which we use to calculate the device WIT score) alongside
inferred two-qubit gate fidelities assuming this simple noise model. 

The mapping $F_{2q}=E^{1/24}$ is intended only as an \emph{effective} two-qubit
fidelity proxy under a simplified noise model in which (i) errors are
approximately gate-local and (ii) the dominant degradation scales with the
number of two-qubit operations. In practice, SPAM, idle errors, leakage, and
coherent effects can contribute substantially. Accordingly, we report $F_{2q}$
as a qualitative normalization to circuit complexity rather than a direct
estimate of any provider-reported native gate fidelity.
Only the measured expectation value enters the WIT subscore 
(normalized in Table~\ref{tab:metriq-summary}) 
and the composite Metriq score; inferred $F_{2q}$ is diagnostic only.

The WIT benchmark demonstrates particular value for assessing device suitability
for quantum simulation applications. Unlike abstract metrics such as Quantum
Volume that condense multiple performance dimensions into a single number, WIT
directly probes the types of operations required for simulating physical
systems: parameterized unitary evolution, time-reversal symmetry, and
information preservation through complex multi-qubit dynamics.

\begin{table}[H]
    \centering
    \begin{tabular}{lcc}
        \toprule
        \textbf{Device} & \textbf{Expectation} & \textbf{Effective 2Q-fidelity proxy $F_{\text{2q}}$}\\
        \midrule
        \texttt{quantinuum\_h2\_2} & 0.981 & 99.9\% \\
        \texttt{ibm\_pittsburgh} & 0.913 & 99.6\% \\
        \texttt{ibm\_boston} & 0.897 & 99.5\% \\
        \texttt{ibm\_brisbane} & 0.499 & 97.2\% \\
        \texttt{ibm\_marrakesh} & 0.875 & 99.4\% \\
        \texttt{ibm\_fez} & 0.865 & 99.4\% \\
        \texttt{ibm\_kingston} & 0.854 & 99.3\% \\
        \texttt{ibm\_torino} & 0.773 & 98.9\% \\
        \texttt{iqm\_garnet} & 0.713 & 98.6\% \\
        \texttt{iqm\_emerald} & 0.665 & 98.3\% \\
        \texttt{rigetti\_ankaa\_3} & 0.463 & 96.8\% \\
        \texttt{wukong\_72} & 0.440 & 96.6\% \\
        \texttt{wukong\_102} & 0.376 & 96.0\% \\
        \bottomrule
    \end{tabular}
    \caption{WIT benchmark performance across tested devices. 
    WIT is evaluated at a single width ($n=7$), so the width aggregation is trivial. 
    Inferred two-qubit gate fidelities are computed as $F_{\text{2q}} = E^{1/24}$,
    where $E$ is the expectation value and 24 is the number of two-qubit gates in the circuit.
    Data captured October--December 2025.}
    \label{tab:wit-performance}
\end{table}

%------------------------------------------------------------------------------%
\subsubsection{Linear-ramp quantum approximate optimization algorithm (LR-QAOA)}
\label{sec:benchmark-qaoa}
%------------------------------------------------------------------------------%

The \emph{linear‑ramp quantum approximate optimization algorithm} (LR‑QAOA) is a variant of the quantum approximate optimization algorithm (QAOA) that
forgoes variational parameter optimization in favor of a predetermined,
layer‑dependent schedule. Standard QAOA uses two sets of variational angles
$\{\gamma_j,\beta_j\}_{j=1}^{p}$ to approximately solve combinatorial
optimization problems by applying alternating phase‑separator and mixer
units to an initial product state.  For the \emph{weighted maximum cut}
problem, one is given a graph $G=(V,E,w)$ with real weights on the edges
and seeks a binary string $x\in\{0,1\}^{|V|}$ that maximizes the cut value
\begin{equation}
    \label{eq:maxcut}
    C(x)=\sum_{(i,j)\in E} w_{ij}[x_i \oplus x_j],
\end{equation}
where $x_i\oplus x_j=1$ whenever the bits $x_i$ and $x_j$ differ.  In QAOA, the
phase‑separator is generated by the weighted cut operator
$\hat{H}_C=\sum_{(i,j)\in E} w_{ij} Z_i Z_j/2$ and the mixer is
$\hat{H}_M=\sum_{i} X_i$ acting on qubits associated with vertices. Instead of
running the variational circuit, where the rotation angles are optimized using a
classical computer, in our implementation, we consider the LR‑QAOA protocol that
fixes the angles to a linear ramp,
\begin{equation}
    \gamma_j=j\,\Delta\gamma/p, \quad \beta_j=(p+1-j)\,\Delta\beta/p
\end{equation}
for
$j=1,\dots,p$, where $p$ is the QAOA layer and $\Delta\gamma,\Delta\beta
\in[0,2]$ are slope parameters. The resulting depth‑$p$ circuit,
\begin{equation}
    \ket{\psi(\boldsymbol{\gamma},\boldsymbol{\beta})} =
    \left[\prod_{j=1}^{p}
    \mathrm{e}^{-\mathrm{i}\beta_j \hat{H}_M}\,\mathrm{e}^{-\mathrm{i}\gamma_j \hat{H}_C}
    \right]
    \ket{+}^{\otimes N},
\end{equation}
does not require any classical optimizer. Recent work has shown that such linear
schedules can approximate optimal QAOA parameters for diverse combinatorial
problems and yield performance improvements as the depth grows, with success
probabilities scaling as $P(x^*)\approx 2^{-\eta(p)N + C}$ with the function
$\eta(p)$ decreasing with $p$~\cite{montanez2025evaluating,
montanez2025toward}.

The \mgym implementation of the LR‑QAOA benchmark samples the performance of a
quantum device on the weighted MaxCut problem under this linear schedule.  The
user specifies a \emph{graph type} (1D chain, native layout or fully connected),
the \emph{number of qubits} \(N\), a list of \emph{qaoa\_layers} \(p\) and
slopes \(\Delta\beta,\Delta\gamma\).  A weighted graph of size \(N\) is
constructed: for \(\mathrm{1D}\) a chain with edges \((i,i+1)\), for
\(\mathrm{NL}\) the connectivity of the target device is used, and for
\(\mathrm{FC}\) all pairs \((i,j)\) are included (an additional SWAP network is
inserted if the device is not fully connected).  Edge weights \(w_{ij}\) are
drawn from a small set (\(\{0.1,0.2,0.3,0.5,1.0\}\)) and the optimal bitstring
\(x^*\) is computed classically with a simulated annealing solver. For each
depth \(p\), the LR‑QAOA circuit with the linear‑ramp parameters above is
executed several times (\texttt{shots}) across multiple trials.  The outcome
distribution \(\{\hat{x}\}\) is used to compute the empirical \emph{approximation ratio}
\begin{equation}
    \label{eq:approxratio}
    \mathrm{r} = \frac{1}{M}\sum_{\hat{x}} \frac{C(\hat{x})}{C(x^*)},
\end{equation}
where \(M\) is the total number of shots.  Additionally, the probability
\(P(\hat{x}=x^*)\) of sampling the optimal solution and a \emph{random} baseline
are estimated by drawing uniformly random bitstrings and computing their average
cut value.  A one‑sided \(t\)-test compares the observed approximation ratio
against the random mean to determine whether the result significantly exceeds
random sampling at a chosen confidence level.  The benchmark returns the list of
approximation ratios that correspond to the list of layers, the random baseline
\(\mathrm{r}_\text{rand}\) and a Boolean flag indicating whether the test passes
at each depth.

LR‑QAOA provides a deterministic, scalable method to probe quantum hardware at
large width and depth.  Unlike variational QAOA, no classical optimization is
required, so circuits can contain hundreds of layers and
qubits~\cite{montanez2025evaluating}. 
By comparing the approximation ratio against a random baseline, we quantify how long a device can
preserve coherent quantum signal as the number of circuit layers increases under the assumption that the asymptotic infinite-layer limit approaches the adiabatic limit.  The ability to
choose different graph topologies (\(\mathrm{1D}\), native layout and fully
connected) allows the benchmark to stress both nearest‑neighbor and all‑to‑all
couplings, complementing constant‑depth and mirror‑circuit benchmarks in our
suite.   Related work has explored pushing QAOA to large depth--width regimes and improving robustness via quantum error detection and co-compilation, providing additional motivation for including scalable QAOA-style workloads such as LR-QAOA in our benchmark suite~\cite{jpm2023qaoa,jpm2025qaoa,jpm2025qaoa2}. As QAOA and its variants have shown evidence of a scaling advantage for certain
combinatorial problems over classical heuristics~\cite{shaydulin2024evidence,
montanez2025toward}, incorporating it into our suite provides insight into when
quantum devices may deliver algorithmic improvements beyond random sampling,
making it a valuable tool for tracking progress in quantum computing.

For LR-QAOA, we summarize device performance across several problem sizes into a
single scalar score. For each target qubit number $N \in \{10,20,50,100\}$ we
run LR-QAOA on a weighted MaxCut instance defined on a 1D chain of $N$ qubits,
using the benchmark configuration of the number of shots per circuit
$\texttt{shots} = 1000$, number of circuit trials $\texttt{trials} = 10$,
with a fixed QAOA layer $p = 10$ and slope parameters $\Delta\beta = 0.3$
and $\Delta\gamma = 0.6$.  For a given instance, we estimate the approximation
ratio $r$.  A random baseline $r_{\mathrm{random}}$ is obtained in the same
way by sampling bitstrings from a uniform distribution over $\{0,1\}^N$ with
$\texttt{num\_random\_trials} = 10$ independent random experiments with $1000$
shots each.  We then define the effective approximation ratio
\begin{equation}
    r_{\mathrm{eff}} = \frac{r - r_{\mathrm{random}}}{1 - r_{\mathrm{random}}},
\end{equation}
so that \(r_{\mathrm{eff}} = 0\) corresponds to random sampling and
\(r_{\mathrm{eff}} = 1\) corresponds to perfectly sampling optimal bitstrings.
The quantity \(\text{LR-QAOA-N}\) reported is \(r_{\mathrm{eff}}\) averaged over
the \(\texttt{trials} = 10\) problem instances at that \(N\). %the confidence
level parameter is used to verify that the observed approximation ratios
significantly exceed the random baseline for each setting. 

The overall LR-QAOA score for a device is obtained by aggregating the effective
approximation ratios across the four problem sizes using weights proportional
to circuit width (Eq.~\eqref{eq:width-weights}). Let
$\text{LR-QAOA-}N$ denote the effective approximation ratio obtained for a
1D chain of $n$ qubits at fixed depth $p=10$. The width-aggregated raw value is
\begin{equation}
    \textbf{LR-QAOA Score}
    =
    \frac{
      10\,\text{LR-QAOA-10} +
      20\,\text{LR-QAOA-20} +
      50\,\text{LR-QAOA-50} +
      100\,\text{LR-QAOA-100}
    }{180}.
\end{equation}
This aggregation emphasizes performance at larger system sizes while still
crediting performance on smaller chains. Devices that can maintain good
approximation ratios as the problem size increases, therefore achieve larger
values of $\text{LR-QAOA Score}$. For devices with fewer than $n$ qubits, the
corresponding term $\text{LR-QAOA-}N$ is set to zero, so devices unable to reach
larger problem sizes are penalized accordingly.

For completeness, the schema used to generate the results (in this case, for a LR-QAOA-10 job) 
is given as
\begin{verbatim}
"benchmark_name": "Linear Ramp QAOA",
"graph_type": "1D",
"num_qubits": 10,
"shots": 1000,
"trials": 10,
"confidence_level": 0.999,
"num_random_trials":25,
"seed":123,
"qaoa_layers":[10],
"delta_beta":0.3,
"delta_gamma":0.6
\end{verbatim}

The results are summarized in Table~\ref{tab:lqaoa-results}.

\begin{table}[H]
    \centering
    \begin{tabular}{lcccccc}
    \toprule
    \textbf{Device} & \textbf{Qubits} & \textbf{$r_\text{eff}(N=10)$} & \textbf{$r_\text{eff}(N=20)$} & \textbf{$r_\text{eff}(N=50)$} & \textbf{$r_\text{eff}(N=100)$} & \textbf{LR-QAOA Score} \\
    \midrule
    \texttt{ibm\_boston}        & 156 & 0.68305 & 0.66408 & 0.68739 & 0.67340 & 0.676787 \\
    \texttt{ibm\_pittsburgh}    & 156 & 0.67055 & 0.62562 & 0.68327 & 0.64690 & 0.655952 \\
    \texttt{ibm\_kingston}      & 156 & 0.66456 & 0.61066 & 0.66866 & 0.61159 & 0.630282 \\
    \texttt{ibm\_marrakesh}     & 156 & 0.60698 & 0.55631 & 0.64030 & 0.57929 & 0.595222 \\
    \texttt{ibm\_fez}           & 156 & 0.57550 & 0.52658 & 0.59774 & 0.46365 & 0.514103 \\
    \texttt{ibm\_torino}        & 133 & 0.56820 & 0.50338 & 0.37939 & 0.35449 & 0.389823 \\
    \texttt{quantinuum\_h2\_2}  & 56  & 0.76358 & 0.74043 & 0.80175 & --      & 0.347399 \\
    \texttt{iqm\_emerald}       & 54  & 0.52483 & 0.50756 & 0.00992 & --      & 0.088308 \\
    \texttt{iqm\_garnet}        & 20  & 0.52881 & 0.14321 & --      & --      & 0.045291 \\
    \texttt{wukong\_72}         & 72  & 0.17308 & 0       & 0       & --      & 0.009616 \\
    \texttt{rigetti\_ankaa\_3}  & 82  & 0.03372 & $0^*$   & $0^*$   & --      & 0.001873 \\
    \bottomrule
    \end{tabular}
    \caption{LR-QAOA effective approximation ratio $r_\mathrm{eff}$ (with QAOA
    layer $p=10$) as well as the LR-QAOA score across devices and qubit counts.
    For \texttt{wukong\_72}, despite the device having
    a sufficient amount of qubits for $N=20$ and $N=50$, due to device issues,
    the benchmark was only able to run for $N=10$ and was unable to run for
    $N=100$ due to insufficient device qubits. Hence, the \texttt{wukong\_72}
    device has a score of $0$ for $N \in \{20, 50, 100\}$.  Data captured November 2025--February 2026.}
    \label{tab:lqaoa-results}
\end{table}

%------------------------------------------------------------------------------%
\subsubsection{Quantum Fourier transform (QFT)}
\label{sec:benchmark-qft}
%------------------------------------------------------------------------------%

This section describes the \textit{Quantum Fourier Transform} (QFT) benchmark within \mgym. Unlike low-level diagnostic benchmarks that measure isolated hardware characteristics (e.g., $T_1$, $T_2$, or individual gate fidelities), the QFT benchmark probes a processor’s capacity to maintain coherence and accumulate phase information throughout a structured, multi-qubit quantum circuit, which are properties essential for a wide range of quantum algorithms.
The QFT is a unitary transformation mapping computational basis states to the quantum Fourier basis,
\begin{equation}
    \mathrm{QFT}\,\lvert x\rangle
    = \frac{1}{\sqrt{2^n}}
      \sum_{k=0}^{2^n-1}
        e^{2\pi i\,xk/2^n}\,\lvert k\rangle.
\end{equation}
Each computational basis state $\lvert x\rangle$ becomes an equal-amplitude superposition with phases encoding $x$. The QFT is a core primitive for algorithms such as Shor’s factoring algorithm, quantum phase estimation, hidden subgroup problems, and numerous quantum signal processing workflows.

The \mgym QFT benchmark is integrated using the QED-C Application-Oriented Performance Benchmarks for Quantum Computing suite~\cite{lubinski2023application}. We support the following variants:
\begin{itemize}
\item Method 1: initializes an $n$-qubit register to a uniformly random computational basis state, applies the QFT, performs increment-by-one modulo $2^n$ in the Fourier basis, and applies the inverse QFT to return to the computational basis. The ideal output is the deterministic state $\lvert x+1 \bmod 2^n \rangle$, making deviations directly attributable to accumulated noise during forward QFT, Fourier-basis arithmetic, and inverse QFT.

\item Method 2: encodes a random integer $x$ directly into the Fourier basis using Hadamard and $Z$-rotation gates, then applies the inverse QFT to recover $\lvert x\rangle$. This approach isolates the inverse QFT and Fourier-basis encoding, providing complementary algorithmic coverage with a shorter circuit.
\end{itemize}

\noindent The benchmark configuration in \mgym mirrors the QED-C specification. An example is below:
\begin{verbatim}
"benchmark_name": "Quantum Fourier Transform",
"shots": 1000,
"min_qubits": 4,
"max_qubits": 20,
"skip_qubits": 4,
"max_circuits": 3,
"method": 1,
"use_midcircuit_measurement": false
\end{verbatim}
This benchmark executes a sweep over problem sizes from \texttt{min\_qubits} to \texttt{max\_qubits} in increments of \texttt{skip\_qubits}. For each size, up to \texttt{max\_circuits} parameterized circuits are generated. The \texttt{method} field selects between the two benchmark variants defined above. The \texttt{use\_midcircuit\_measurement} flag enables an alternative dynamic-circuit implementation of QFT, which can reduce circuit depth by replacing controlled rotations with classically conditioned single-qubit gates~\cite{patel2025platform}.
For each circuit in the sweep, the benchmark computes a normalized fidelity metric~\cite{lubinski2023application} quantifying agreement between the observed and ideal output distributions. \mgym aggregates these fidelities across circuit instances and qubit sizes to produce a single benchmark score. As a result, the QFT benchmark captures the cumulative impact of gate errors, crosstalk, decoherence, and measurement noise across structured multi-qubit operations.

The QFT benchmark is broadly representative of algorithmic workloads. Many quantum algorithms require the precise accumulation of controlled-phase rotations, long-range entangling patterns, and coherent interference between computational paths, all behaviors exercised directly by the QFT. Consequently, performance on the QFT benchmark serves as an indicator of a device’s readiness for practical quantum algorithms in areas such as metrology, chemistry simulation, phase estimation, and cryptographically relevant computations.

\begin{table}
\centering
    \begin{tabular}{lcccccc}
    \toprule
    \textbf{Device} & \textbf{Qubits} & \textbf{QFT-4} & \textbf{QFT-8} & \textbf{QFT-12} & \textbf{QFT-20} & \textbf{QFT Score} \\
    \midrule
    \texttt{quantinuum\_h2\_2}  & 56  & 0.991 & 0.973 & 0.937 & 0.152 & 0.5916 \\
    \texttt{ibm\_boston}        & 156 & 0.862 & 0.420 & 0.038 & 0.000 & 0.1651 \\
    \texttt{ibm\_pittsburgh}    & 156 & 0.828 & 0.368 & 0.004 & 0.000 & 0.1433 \\
    \texttt{ibm\_kingston}      & 156 & 0.845 & 0.154 & 0.002 & 0.000 & 0.1054 \\
    \texttt{ibm\_torino}        & 133 & 0.808 & 0.215 & 0.002 & 0.000 & 0.1131 \\
    \texttt{ibm\_marrakesh}     & 156 & 0.717 & 0.136 & 0.003 & 0.000 & 0.0907 \\
    \texttt{ibm\_fez}           & 156 & 0.674 & 0.024 & 0.001 & 0.000 & 0.0659 \\
    \texttt{iqm\_garnet}        & 20  & 0.495 & 0.027 & 0.000 & 0.000 & 0.0499 \\
    \texttt{iqm\_emerald}       & 54  & 0.384 & 0.002 & 0.001 & 0.000 & 0.0355 \\
    \texttt{rigetti\_ankaa\_3}  & 82  & 0.056 & 0.000 & 0.000 & 0.000 & 0.0051 \\
    \texttt{wukong\_72}         & 72  & 0.016 & 0.000 & 0.001 & 0.000 & 0.0017 \\
    \bottomrule
    \end{tabular}
    \caption{QFT score (Data captured October 2025--February 2026).}
    \label{tab:qft-performance}
\end{table}

We next present the experimental results on the benchmark.
We evaluate QFT at widths $n\in\{4,8,12,20\}$. Let $\text{QFT-N}$ denote
the raw QFT metric at width $n$. We aggregate across widths as
\begin{equation}
  \textbf{QFT Score}
  := \sum_{n\in\{4,8,12,20\}} \frac{n}{44}\, \text{QFT-N}
  \label{eq:qft-width-aggregate}
\end{equation}
(i.e., weights $(1,2,3,5)/11$). The final scores are presented in Table~\ref{tab:qft-performance}.

In addition to QFT, \mgym stages integrations with other QED-C application-oriented benchmarks that can be tested and considered for future inclusion in the \mgym suite. \textit{Bernstein–Vazirani} tests the
ability to implement oracle-based algorithms and maintain coherence through shallow, highly parallel circuits. \textit{Quantum phase estimation} probes long-range phase coherence and controlled-unitary
operations, representative of algorithms in quantum chemistry and Hamiltonian simulation. \textit{Hidden shift} evaluates multi-qubit interference patterns and oracle structure. These benchmarks collectively expand the suite’s future coverage of algorithmic primitives beyond QFT, providing a broader view of device performance across representative computational tasks.

%------------------------------------------------------------------------------%
\section{Benchmarking Costs}
\label{sec:cost-estimation}
%------------------------------------------------------------------------------%
A practical barrier to comprehensive quantum benchmarking is the monetary cost
of executing circuits on commercial quantum hardware. As such, a core principle
guiding \mgym development is frugality, which ensures that benchmarks remain
accessible to the broader quantum computing community, including academic
researchers and smaller organizations with limited budgets. Different providers
employ varying pricing models based on factors such as circuit depth, gate
counts, shot numbers, and device capabilities. Without prior knowledge of these
costs, researchers risk exhausting allocated budgets on exploratory runs or
inadvertently submitting expensive jobs. Given this challenge of estimating
costs across providers ahead of time, in Table~\ref{tab:cost-estimates}, we
report the cost to gather the benchmark results presented in
Section~\ref{sec:benchmarking-suite}. This supplements the primary performance
view with an economic view, which we hope will guide others looking to reproduce
these results, benchmark new devices, and add new benchmarks. All reported cost
metrics correspond to post-run accounting based on provider definitions and
corresponding pricing documentation. Note that how these costs translate to
billed amounts depends on the terms of an organization’s contract with the
provider.

\begin{table}[H]
    \centering
    \begin{tabular}{lcccc}
        \toprule
         & \multicolumn{2}{c}{AWS} & \multicolumn{1}{c}{Quantinuum} & IBM \\
         & \multicolumn{2}{c}{\texttt{iqm\_emerald}} & \multicolumn{1}{c}{\texttt{quantinuum\_h2\_2}} & \texttt{ibm\_torino} \\
        \cmidrule(lr){2-3} \cmidrule(lr){3-4} \cmidrule(lr){5-5}
        \textbf{Benchmark} & Tasks & Shots & HQC & Runtime (s) \\
        \midrule
        BSEQ     & 16 & 16{,}000 &  7{,}625  &  6.05 \\
        CLOPS    & -- &  --  &  -- & 3.25 \\
        EPLG &  60 &  30{,}000    &     17{,}826   &    72.00 \\
        LR-QAOA (50-qubit)  & 10  & 10{,}000  &   11{,}950       & 4.91  \\
        Mirror Circuits (8-qubits, 64-layers) & 10  &  10{,}000    &   7{,}338    & 3.59 \\ 
        QML Kernel (10-qubit) & 1  & 1{,}000     &   109    & 1.14    \\
        QML Kernel (50-qubit) & 1  &   1{,}000   &   565    &   1.27 \\
        QFT (12-qubit) & 3  & 3{,}000  &   1{,}435    &   1.92 \\
        WIT (7-qubit) & 1  &  8{,}192    &   504    &   2.92 \\
        \bottomrule
    \end{tabular}
    \caption{Provider-specific cost metrics for each benchmark in the \mgym suite,
    reported for one representative device per cloud provider: AWS
    (\texttt{iqm\_emerald}), Quantinuum (\texttt{quantinuum\_h2\_2}), and IBM
    (\texttt{ibm\_torino}). Each column reports the quantity relevant to that
    provider's pricing model, task and shot counts for AWS, Hardware Quantum
    Credits (HQC) for Quantinuum, and device runtime for IBM, enabling users to
    estimate monetary costs using the provider's published pricing documentation.}   
    \label{tab:cost-estimates}
\end{table}

In addition to the measured costs reported above, \mgym provides a lightweight
cost estimation utility (\texttt{mgym job estimate}) to support experiment
planning. The tool reports circuit-level statistics prior to execution, allowing
users to identify potentially expensive benchmark configurations before
submitting jobs to hardware. This is particularly useful given that many providers do not
offer pre-run cost estimation APIs.
Because providers employ heterogeneous pricing models, \mgym does not compute
monetary costs directly for all platforms. Instead, it reports statistics such as
gate counts, measurement counts, shot counts, and task counts, which can be mapped
to provider pricing documentation. For some providers, such as Quantinuum, these
statistics translate directly into standardized usage metrics (e.g., Hardware
Quantum Credits), while others, such as Amazon Braket, require manual calculation
based on shots and tasks~\cite{amazon2025pricing}. Estimated costs are therefore
approximate and may differ from realized costs due to provider-side compilation,
execution details, and organization-specific contract terms. As provider pricing models and access agreements continue to evolve, the
estimation framework can be extended to incorporate additional metrics and
updated formulas while maintaining its role as a practical budgeting tool.

%------------------------------------------------------------------------------%
\section{Platform-enabled cross-benchmark analyses}
\label{sec:cross-benchmark}
%------------------------------------------------------------------------------%

\begin{figure}[t]
\centering
\includegraphics[width=1\linewidth]{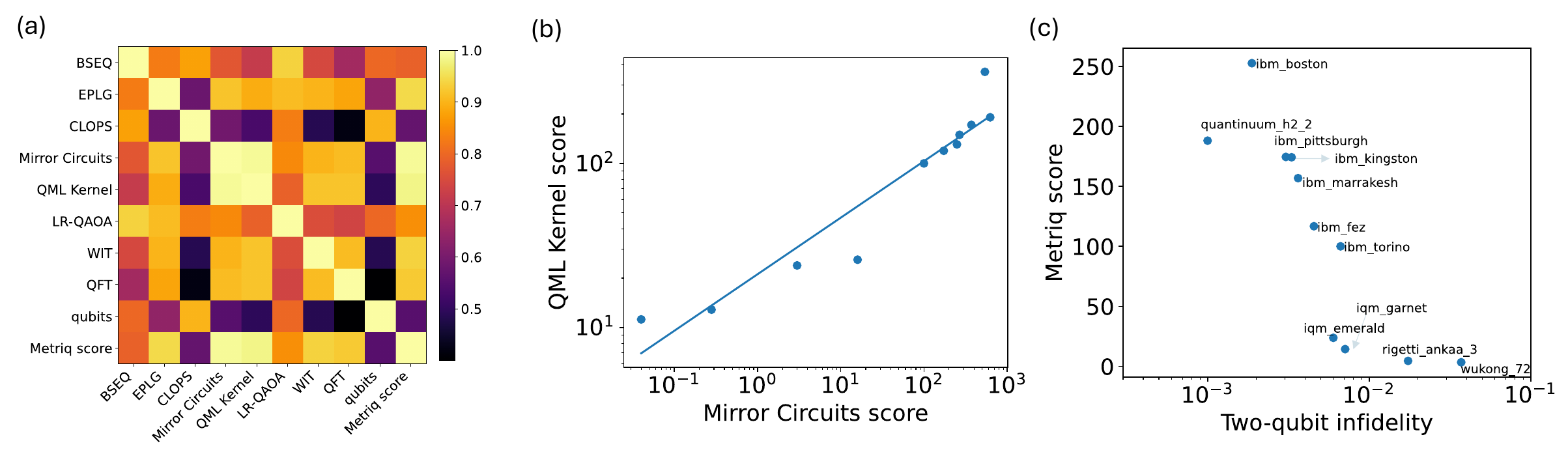}
\caption{Cross-benchmark relationships from Table~\ref{tab:metriq-summary}.
(a) Spearman rank correlations across benchmark component scores, device qubit numbers and the aggregate Metriq score (pairwise-complete
observations). Sample size for the evaluation follows the number of entries with successful benchmark run in Table~\ref{tab:metriq-summary}.
(b) An example proxy relationship showing that Mirror Circuits closely tracks QML Kernel across devices on log scales,
consistent with both benchmarks probing echo-style circuit fidelity under accumulated coherent and two-qubit errors. (c) Correlations between Metriq score and two-qubit infidelity. }
\label{fig:cross_benchmark}
\end{figure}

A central advantage of Metriq is that standardized execution and schema-validated records enable quantitative
\emph{cross-benchmark} insights, rather than isolated per-benchmark reporting. With the results in Table~\ref{tab:metriq-summary},
we compute Spearman rank correlations $\rho$ across the benchmark components and the aggregate Metriq score
(Fig.~\ref{fig:cross_benchmark}(a)). Except for CLOPS that characterizes the speed the circuits, all other pairwise benchmark correlations are positive and large
($0.664 \le \rho \le 0.991$ off-diagonal) as the benchmarks characterize the quality and scale and various circuits. Consistent with this
picture, a PCA of z-scored log-scores (devices with complete component data) yields a first principal component that
explains $88\%$ of the variance.

Several particularly informative couplings emerge. Mirror Circuits and the QML Kernel benchmark are nearly
rank-identical with $\rho=0.991$. This is expected because both are Loschmidt echo-type tests with a deterministic ideal
outcome: the circuit applies a structured unitary followed by an explicit inverse, and the score is
effectively a survival probability that is strongly sensitive to accumulated coherent error, two-qubit gate error, and
compilation-induced depth. In the same spirit, QML Kernel and QFT are also strongly coupled ($\rho=0.918$).
Although they represent different applications, both require coherent phase accumulation and subsequent refocusing
into a known computational-basis outcome (QFT followed by inverse QFT, with an intermediate Fourier-basis
arithmetic primitive), so dephasing and coherent phase noise degrade both benchmarks in a similar, largely monotone
manner across devices. In contrast, BSEQ exhibits its strongest association with LR-QAOA ($\rho=0.936$),
consistent with LR-QAOA sensitivity to entangling capability over the device coupling graph. Also, the BSEQ benchmark is strongly correlated with qubit number as it is defined in terms of the largest connected component of the device graph (Eq.~\ref{eq:bseq_definition}). 
We also find that the aggregated Metriq score is most tightly correlated with Mirror Circuits benchmark ($\rho=0.991$), a benchmark that has been verified both theoretically and experimentally~\cite{proctor2021measuring, proctor2022scalable}.

To test whether the application-inspired benchmark outcomes can be predicted from the tested system-level
benchmarks, we fit a log-linear ridge model using the three system metrics (BSEQ, EPLG, Mirror Circuits) and evaluate
generalization using leave-one-device-out cross-validation.  The resulting cross-validated coefficients of determination on the log scale are $R^2_{\log}=0.925$ for QML (with sample size $n=10$ due to missing EPLG on one
device), $R^2_{\log}=0.922$ for LR-QAOA ($n=10$). For certain workloads, single-metric proxies are particularly stable:
Mirror Circuits alone predicts QML with $R^2_{\log}=0.941$ ($n=11$) as shown in Fig.~\ref{fig:cross_benchmark}(b), while BSEQ alone predicts LR-QAOA with $R^2_{\log}=0.921$ ($n=11$). 

We additionally compare benchmark outcomes to vendor-reported public device calibration summaries (from the relevant vendor websites as of date around March 03, 2026), including two-qubit gate fidelities and readout error where available, and speed metrics such as CLOPS. Across all devices in Table~\ref{tab:metriq-summary} for which two-qubit fidelities are calibrated, the Metriq score is strongly correlated with two-qubit gate fidelity (Spearman $\rho=0.982$), with similar tight coupling for the interference-based Mirror Circuits ($\rho=0.973$) and QML Kernel ($\rho=0.982$). Equivalently, scores exhibit a strong monotone anti-correlation with two-qubit gate error, consistent with multi-qubit gate imperfections being a primary limiter for the circuit depths probed by this suite. An example is shown in Fig.~\ref{fig:cross_benchmark}(c). These comparisons should be interpreted cautiously because calibration metrics are not time-matched to the Metriq runs, and reported fidelities can depend on the gateset and characterization protocol across hardware platforms. In summary, the presented results above illustrate how Metriq enables ecosystem-level insight:
correlations reveal shared physical bottlenecks across benchmark families. At
the same time, deviations from proxy relations provide a quantitative way to
study workload-specific strengths and weaknesses for follow-up work. The
analysis presented here could thus serve as motivation for more follow-up
analysis through the curated dataset.

%------------------------------------------------------------------------------%
\section{Discussions and conclusion}
\label{sec:discussion}
%------------------------------------------------------------------------------%

Several limitations constrain the current scope of \mgym. First, the suite
remains incomplete; many proposed benchmarks have not yet been implemented due
to resource constraints or technical challenges in ensuring cross-platform
compatibility. Second, the cost of executing benchmarks on commercial quantum
hardware remains a practical barrier to comprehensive coverage. While our
frugality principle mitigates this to some extent, exhaustive benchmarking
across all available devices at high frequency is not yet feasible. Third, as
quantum hardware continues to evolve rapidly, particularly with the emergence
of error-corrected logical qubits, our benchmarking methods must adapt
accordingly. The current suite focuses primarily on physical qubits in
gate-based quantum computers, and extension to analog quantum devices and more
sophisticated logical qubit benchmarks represents important future work. 
In this section, we present how quantum error mitigation and circuit compilation impact benchmark 
results and discuss future implementation of logical-level benchmarking.

%------------------------------------------------------------------------------%
\subsection{Quantum error mitigation in benchmarking}
\label{sec:benchmarking-qem}
%------------------------------------------------------------------------------%

Current quantum devices operate in the noisy intermediate-scale quantum (NISQ)
era, where computational results are degraded by gate errors, decoherence,
readout imperfections, and crosstalk. While benchmarking protocols measure
device performance under these realistic conditions, quantum error mitigation
(QEM) techniques offer a complementary perspective by applying algorithmic
corrections to mitigate errors without requiring full quantum error correction
overhead~\cite{cai2023quantum, endo2021hybrid}. Understanding how QEM techniques
interact with benchmark results provides valuable insights into both the
practical utility of mitigation strategies and the fundamental limitations of
current hardware~\cite{li2017efficient, temme2017error,
kandala2019error,koczor2021exponential, strikis2021learning,
giurgica2020digital, endo2018practical, kim2023scalable, he2020zero,
huggins2021virtual, pascuzzi2022computationally, russo2023testing,
song2019quantum}.

The \mgym framework is designed to accommodate optional QEM integration through
the Mitiq software package~\cite{larose2022mitiq}, an open source Python library
implementing a variety of error mitigation protocols. Mitiq provides a
hardware-agnostic interface for applying techniques such as zero-noise
extrapolation (ZNE)~\cite{temme2017error, kandala2019error, giurgica2020digital,
he2020zero, pascuzzi2022computationally, mari2021extending}, probabilistic error
cancellation (PEC)~\cite{mari2021extending, ma2024limitations,
van2023probabilistic}, dynamical decoupling (DD)~\cite{santos2005dynamical,
pokharel2018demonstration, sekatski2016dynamical}, Clifford data regression
(CDR)~\cite{czarnik2021error}, and layerwise Richardson extrapolation
(LRE)~\cite{russo2024quantum} among others. By integrating Mitiq capabilities
into the benchmarking workflow, users could optionally apply one or more
mitigation strategies to benchmark circuits and compare the mitigated results
against raw hardware outputs.

The proposed integration would allow selecting an error mitigation protocol,
which would be applied during circuit execution, collect both raw and mitigated
results, and report both sets of metrics. For techniques like ZNE that require
multiple circuit executions at different noise levels, \mgym would automatically
manage the additional circuit submissions and aggregate the extrapolated
results. For PEC, which requires characterizing the noise model through gate
fidelity measurements, the framework could either use cached calibration data or
perform on-the-fly characterization depending on data availability and user
preferences.

However, implementing such a unified QEM workflow across heterogeneous cloud
platforms presents significant practical challenges. Provider-specific
constraints on job submission complicate automated mitigation: for instance,
AWS Braket requires submitting each circuit as a separate job, substantially
increasing the overhead for techniques like ZNE that require executing multiple
noise-scaled variants per benchmark circuit. In contrast, IBM Quantum and IonQ
platforms allow batch submissions that can more efficiently accommodate the
additional circuits needed for mitigation protocols. These platform differences
mean that a single QEM-enabled benchmark suite cannot be deployed uniformly;
instead, the framework would need provider-aware scheduling logic to optimize
circuit submission patterns while maintaining equivalent mitigation protocols
across platforms.

The value of QEM-enhanced benchmarking extends beyond simply reporting improved
performance metrics. First, comparing raw and mitigated results across different
devices reveals which platforms benefit most from particular mitigation
strategies, providing actionable guidance for algorithm deployment. 
Second, tracking the \emph{mitigation efficiency}, that is, the ratio of mitigated to
raw performance, serves as a meta-metric characterizing how amenable a device is
to algorithmic error suppression. Devices with high mitigation efficiency may
achieve practical utility sooner than devices with superior raw performance but
poor mitigation response.
Third, QEM integration enables more nuanced interpretation of benchmark results
in the context of near-term quantum advantage. Many quantum algorithms proposed
for NISQ-era applications implicitly assume some level of error mitigation will
be applied during execution. By reporting both raw and mitigated benchmark
scores, \mgym would provide a more realistic assessment of achievable performance for
practical workloads. For example, the WIT benchmark's demanding coherence
requirements might yield expectation values near 50\% on raw hardware but exceed
80\% with ZNE, substantially altering conclusions about the device's suitability
for quantum simulation applications.

However, QEM integration also introduces methodological considerations that must
be addressed carefully. Different mitigation techniques make different
assumptions about noise structure and may perform unpredictably when those
assumptions are violated~\cite{govia2025bounding}. PEC %assumes gate-independent Markovian noise, 
needs Markovian noise for each circuit layer and the single qubits to have gate-independent error, which breaks down in the presence of strong crosstalk or time-correlated errors. ZNE
requires that noise scales consistently under circuit modifications, which may
not hold for all error sources. Reporting mitigated results without clearly
documenting the applied techniques and their assumptions risks obscuring rather
than illuminating device capabilities.

Furthermore, the sampling overhead of some QEM techniques can be substantial.
PEC's requirement for thousands of additional shots may render it impractical
for cost-constrained benchmarking, particularly on providers charging per shot.
Even ZNE's relatively modest overhead (typically $2$--$5\times$ additional
shots) accumulates quickly when applied across comprehensive benchmark suites.
The \mgym cost estimation feature discussed in Section~\ref{sec:cost-estimation}
becomes even more critical in the QEM context. Cost structures vary dramatically
across providers: trapped-ion platforms (IonQ, Quantinuum) typically charge per
circuit or gate count, while superconducting platforms (IBM, Rigetti) may offer
free access tiers but with limited queue priority. ZNE's $2$--$5\times$ overhead
multiplies these baseline costs accordingly, and PEC's requirement for thousands
of additional shots can render it prohibitively expensive on per-shot billing
models. Users must weigh the value of mitigated results against
platform-specific resource consumption, with the cost estimation framework
providing quantitative guidance for these trade-offs before committing to
QEM-enhanced benchmark runs.

Looking forward, QEM-enhanced benchmarking represents a natural evolution of the
\mgym platform as quantum hardware matures and mitigation techniques become
standard practice. The modular design of both \mgym and Mitiq facilitates this
integration: benchmark definitions remain unchanged, while an optional
mitigation layer intercepts circuit execution to apply selected techniques
transparently. Initial integration efforts could focus on the most mature and
widely applicable methods (e.g., ZNE, readout error mitigation), with more
specialized techniques (e.g., PEC, CDR) added as community demand and
implementation maturity warrant. By maintaining QEM as an optional enhancement
rather than a mandatory component, \mgym preserves the ability to report both
raw and mitigated performance, supporting diverse use cases from fundamental
device characterization to applied algorithm development. As providers
standardize their interfaces and mitigation becomes more tightly integrated at
the cloud platform level, the practical barriers to cross-platform QEM
benchmarking will diminish, enabling more direct comparisons of mitigation
efficacy across diverse hardware architectures.

%------------------------------------------------------------------------------%
\subsection{Quantum circuit compilation in benchmarking}
\label{sec:benchmarking-compilation}
%------------------------------------------------------------------------------%

Proper quantum circuit compilation (also referred to as \emph{transpilation}) is an indispensable component of any benchmarking protocol.  Compilation bridges the gap between high-level algorithmic descriptions and the native gate sets, connectivity and timing constraints of a quantum processor.  When circuits are compiled intelligently, they exploit device-specific optimizations such as qubit mapping, gate cancellations and pulse-level calibrations, leading to shorter depths and higher fidelities.  Conversely, naive or inappropriate compilation can either conceal hardware deficiencies by over-simplifying circuits, or inadvertently degrade performance by introducing unnecessary overhead.  Thus, careful control over compilation is essential for fair comparisons across devices and for drawing meaningful conclusions from benchmarking results.

First, in the context of software benchmarking, compiler quality itself becomes the subject of investigation~\cite{nation2024benchmarking}.  
%For instance, the \texttt{ucc\_bench} project evaluates unitary coupled-cluster circuits across different transpilers to assess how effectively each compiler reduces depth and two-qubit gate counts on superconducting hardware~\cite{https://github.com/unitaryfoundation/ucc-bench}.  
Studies assessing how effectively each compiler reduces depth and optimizes two-qubit gate counts~\cite{nation2024benchmarking,quetschlich2023mqt,unitary2025ucc,kharkov2022arline}  reveal substantial differences between compilation strategies and underscore the need to benchmark software stacks alongside hardware.
Secondly, compilation is equally critical in hardware benchmarking.  Our QML kernel benchmark (Section~\ref{sec:benchmark-qml}) highlights this point.  The benchmark circuit consists of two symmetric halves separated by a barrier, and the barrier prevents the transpiler from commuting gates across the midpoint and thereby preserves the circuit’s symmetry.  When we executed this circuit using Qiskit’s default transpiler, the barrier ensured that each half was compiled separately and then joined, resulting in the expected inner-product circuit.  However, early tests on the AWS Braket revealed a failure as barriers were not supported on Braket (at the time of data collection), so the transpiler freely canceled gates across the midpoint and reduced the entire circuit to the identity, making the measured accuracy close to unity completely obscuring the hardware's noise characteristics.  This incident demonstrates that if compilation details are not handled correctly, benchmarking results can be rendered meaningless.

To remedy this issue, we adopted the \emph{verbatim compilation} feature provided by AWS Braket~\cite{amazon2025braket}.  Verbatim compilation bypasses the optimizer entirely: the user supplies a circuit expressed in the device’s native gate set, and the device executes it exactly as written, without any rewiring or gate synthesis.  In our case, we first built the QML kernel circuit in Qiskit with the necessary barrier, transposed it into the native gates of the target device (e.g.\ `rz', `rx' and `iswap' for Rigetti device) using Qiskit’s transpiler, converted the result into a Braket circuit via the \texttt{qbraid} toolkit, and then encapsulated it inside a \texttt{verbatim\_box} before submission.  Executing the verbatim circuit on IQM’s \texttt{iqm\_emerald} QPU produced sensible results: for example, for a 20-qubit kernel benchmark the compiled circuit contained 76 two-qubit entangling gates, matching the two-qubit gate count of the underlying Qiskit circuit, and the measured accuracies followed the trends seen on IBM hardware.  Without verbatim compilation, the Braket transpiler would have attempted to optimize away the symmetric halves of the circuit, leading to a trivial identity operation.  Similar workflows were applied to Rigetti and other IQM devices for which we run through AWS Braket, each with their own native gate sets, to ensure fair comparisons. Additional details can be found in Appendix~\ref{app:verbatim_compilation}.

This example underscores several key insights.  First, benchmarking frameworks must explicitly document compilation settings and ensure that barriers or other structural constraints are respected; otherwise, cross-platform comparisons can be misleading.  Second, access to verbatim compilation or pulse-level control is invaluable for hardware benchmarking because it allows researchers to run circuits exactly as designed and to decouple hardware performance from compiler heuristics.  Finally, the interplay between hardware and software layers means that benchmarking should be viewed holistically: only by considering both the quantum processor and the compilation stack can one obtain a realistic picture of near-term quantum capabilities.

%------------------------------------------------------------------------------%
\subsection{Benchmarking fault-tolerant quantum computers}
\label{sec:benchmarking-fault-tolerance}
%------------------------------------------------------------------------------%
As quantum processors mature, conventional benchmarks based on physical qubit error rates and gate fidelities become insufficient to predict the performance
of fault-tolerant computation.  When quantum error correction (QEC) is employed,
operations take place in the \emph{logical} space of an encoded qubit, and the
relevant performance metrics are determined by the interplay of many physical
errors, syndrome extraction, decoder \emph{throughput} and \emph{latency}, and the
architecture of the code.
Recent experiments have demonstrated entangled logical qubits on superconducting
devices using surface-code and Bacon–Shor encodings~\cite{hetenyi2024creating},
long-lived logical memories on heavy-hex superconducting platforms and
trapped-ion systems~\cite{google2025quantum, paetznick2024demonstration}, logical
teleportation on trapped-ion processors~\cite{ryananderson2024high}, and
magic-state preparation beyond distillation
thresholds~\cite{sales2025experimental, ye2023logical, kim2024magic}.  These milestones
signal that logical qubits have advanced from a theoretical construct to an
experimental reality, motivating systematic benchmarking at the logical level.

Benchmarks for logical qubits must assess both the quality of encoded states and the reliability of encoded operations.  Unlike physical benchmarking, where gate infidelity and decoherence are measured on individual qubits, logical benchmarking evaluates the entire QEC stack: state preparation, syndrome extraction cycles, decoding, and logical gates.  Logical error rates depend on correlations across physical qubits and the efficiency of decoding, so they cannot be inferred directly from physical gate fidelities.  Key metrics include the logical memory lifetime (measured in rounds of error correction), average error per logical Clifford gate (via logical randomized benchmarking), the logical gate speed (e.g., logical cycle time or time per logical gate, including the required syndrome-extraction rounds), and the fidelity of resource states such as Bell pairs or magic states.  In addition to reporting absolute logical error rates, it is often informative to report \emph{scaling} with code distance, for example via an error-suppression factor
%(sometimes summarized as a ``$\lambda$'' parameter)
that quantifies how increasing distance improves logical error suppression~\cite{google2025quantum}.  Recent demonstrations have reported logical error rates far below the corresponding physical error rates on trapped-ion processors~\cite{paetznick2024demonstration}
and have observed entanglement fidelities of $\sim94\%$ between encoded surface-code and Bacon–Shor qubits on IBM heavy-hex devices~\cite{hetenyi2024creating}. In neutral-atom processors, encoded qubits have enabled algorithms and loss correction with better-than-physical error rates~\cite{bluvstein2024logical}.

Different levels of logical benchmarking can be organized hierarchically. We remark that this hierarchy is a useful taxonomy for organizing experiments and reporting results, but it should not be interpreted as implying that performance is necessarily composable across levels: logical errors can be non-Markovian and spatially correlated across many qubits and many rounds, because syndrome extraction repeatedly couples the code to ancillas, leakage and coherent errors can persist across cycles, and the decoder closes a feedback loop through Pauli-frame updates and feedforward. As a result, component-level figures of merit may not reliably predict system-level or application-inspired behavior, so meaningful evaluation typically requires running benchmarks at multiple levels.
%  Logical randomized benchmarking protocols provide a device-independent way to estimate the average error rate of logical gates and the probabilities of correctable and uncorrectable errors, capturing crosstalk and correlated error effects not visible at the physical level~\cite{combes2017logical}.
At the \emph{component level}, one benchmarks individual logical operations:
logical Clifford gates via logical randomized benchmarking~\cite{combes2017logical},
non-Clifford gates and magic-state preparation~\cite{dasu2025breaking}, and
ancilla factories or syndrome extraction circuits.  At the \emph{system level},
benchmarks assess multi-logical-qubit primitives such as logical Bell-pair
factories~\cite{paetznick2024demonstration}, logical
teleportation~\cite{bluvstein2025fault,ryananderson2024high}, logical GHZ state
preparation~\cite{hong2024entangling}, and algorithmic kernels such as phase
estimation. Finally, \emph{application-inspired} benchmarks run small
fault-tolerant algorithms end-to-end to measure cumulative logical error rates,
decoding latency and resource overhead.

%\CL{to modify after implementation.}
As a first step towards logical benchmarking in \mgym, we propose the
implementation of a system-level benchmark in the framework: a \emph{logical
Bell-pair factory}. In this benchmark, two distance-three surface-code patches
(on a superconducting device in an initial implementation) are prepared in $|0\rangle$ states,
entangled via a \emph{configurable logical two-qubit entangling primitive} (e.g.,
lattice surgery for patch-based surface codes, or a transversal logical two-qubit
gate when available), and measured in the logical $X$ and $Z$ bases.
Corrections are handled via Pauli\-frame tracking, and post-selection on syndrome outcomes
yields an estimate of the entanglement fidelity $F_{\mathrm{LB}}$ and the
success probability $p_{\mathrm{succ}}$.  By setting the code family, distance,
and the logical entangling method (e.g., lattice surgery vs.\ transversal) as
parameters, \mgym would enable fairer comparisons across hardware platforms
(including superconducting, trapped-ion, and neutral-atom systems) and across
encodings. The logical Bell-pair
factory captures system-level plumbing and routing overhead while remaining
simple to implement on real devices.

The outlook for logical benchmarking is rich and challenging. As physical qubit
counts grow and devices acquire dynamic-circuit capabilities, benchmarks must
incorporate feedforward operations and dynamic lattices and benchmark techniques for these capabilities are beginning to emerge~\cite{eickbusch2025dynamic,govia2023randomized,shirizly2025randomized,hothem2025measuring,zhang2025generalized}. 
Magic-state factories will be critical to universal fault-tolerant computation; benchmarking their fidelity and throughput in realistic noise environments is an urgent next step. Cross-platform comparisons will require standardizing QEC codes and decoding interfaces, because surface, color, and Bacon–Shor codes have different resource trade-offs and gate schedules.  Hardware constraints such as sparse connectivity on superconducting chips, finite measurement latencies, and limited qubit parallelism impose scheduling challenges for lattice surgery and transversal gates~\cite{gutierrez2019transversality}. Decoders must satisfy both
\emph{throughput} (syndromes decoded per unit time) and \emph{latency} (time from
syndrome availability to a correction or Pauli-frame update): insufficient
throughput leads to a growing backlog of unprocessed syndromes that can halt
real-time operation, while latency contributes directly to algorithm wall-clock
time when the quantum control stack must wait on decoder output. Google has
demonstrated real-time decoding on a distance-seven surface-code memory with
63\,$\mu s$ latency~\cite{google2025quantum}, illustrating the importance of co\-designing benchmarks with control electronics. Moreover, the decoding landscape is rapidly evolving~\cite{demartiiolius2024decoding,bausch2024learning}, with a proliferation of algorithmic approaches (and hardware-accelerated implementations) that trade off logical error rate against latency, throughput,
and implementation complexity; logical benchmarks should therefore report the
decoder family and operating point alongside logical performance metrics.
Future benchmarks should gradually incorporate dynamic circuit capabilities,
state-dependent branching, and intermediate resets, enabling a full evaluation
of fault-tolerant protocols such as teleportation and magic-state injection. The output would be the logical
benchmark results for different platforms with various supported QEC codes such
as surface codes, color codes, or quantum Low-Density Parity-Check codes.
Ultimately, benchmarking logical qubits will guide the community toward
architectures and encodings that deliver long\-term stability, high-fidelity
logical gates and resource efficiency on the path to quantum
utility~\cite{proctor2025benchmarking}.

%------------------------------------------------------------------------------%
\subsection{Conclusion and outlook}
%------------------------------------------------------------------------------%

In summary, Metriq provides a unified, open, and continuously evolving platform for  benchmarking quantum computers across heterogeneous hardware providers. By  coupling a declarative benchmark runner, a versioned and schema-validated 
dataset, and an interactive web portal, Metriq enables reproducible and  longitudinal evaluation of quantum devices that is not achievable with  vendor-specific or one-off benchmarking studies. The framework supports both
system-level and application-inspired benchmarks with cost estimations, and the benchmark suite presented here includes both established methods and new benchmarks such as BSEQ executed on hardware platforms where they have not previously been reported. 
The public dataset and the composite Metriq score not only
translate these results into transparent, comparable indicators of hardware capability that can be revisited as devices evolve, but also provide physical insights on the strengths and weaknesses of the benchmark methods.
The living nature of the platform is particularly important: benchmarks can be  periodically re-executed using identical specifications, enabling quantitative  tracking of hardware improvements, calibration changes, and architectural 
updates. This dynamic perspective complements traditional point-in-time  publications and provides stakeholders with up-to-date empirical evidence about  the state of quantum technologies. This approach is in accordance with the FAIR principles, enabling reuse thanks to the machine-readable nature of the digital assets. 

Several directions naturally extend from this work. As discussed above, as early fault-tolerant devices emerge, logical-qubit benchmarks and logical-level performance models
will become essential, and Metriq’s schema-based design provides a structured pathway for incorporating them. Expanding support to additional cloud services,
laboratory systems, and high-performance simulators will further broaden the scope of cross-platform assessments. In particular, it will be important to include platforms that are not as readily available via cloud access or require adapted compilation to run benchmarks, such as neutral atom-based, photonics-based, and spin-based systems, as well as integrating into systemic benchmarks devices based on bosonic or measurement-based quantum computing. Research laboratories in academia, as well as HPC centers and national laboratories, at the institutional level, could adopt \mgym and populate \mweb with results even when they do not provide cloud access, thanks to the portability of the tool. In this context, \mgym could be extended to further integrate HPC and quantum computing resources for HPC-QC benchmarking \cite{chichereau2025hpcqcmark}. Finally, enhanced community features, including programmatic APIs, annotations, and discussion channels may transform Metriq into a collaborative hub for benchmarking research.

Overall, Metriq establishes an independent and extensible foundation for tracking progress in quantum computing. By enabling transparent, reproducible,
and continuous benchmarking across providers and modalities, it supports a
shared empirical basis for evaluating quantum hardware and contributes to the
broader development of reliable and scalable quantum technologies.

\noindent\textbf{Disclaimer.} The hardware results reported in this work were collected at different times (from March 2025 to February 2026) and can be  compilation- or settings-dependent. Accordingly, the benchmarking scores are intended for reference only and should not be interpreted as a definitive or fully up-to-date ranking of device performance.
%------------------------------------------------------------------------------%
\section*{Acknowledgments}
%------------------------------------------------------------------------------%

We thank all early contributors to the Metriq platform repositories for their code contributions and
feedback during the development of the framework. In particular, we acknowledge Alejandro Montañez-Barrera for assisting with the porting of the LR-QAOA benchmark implementation into \mgym.
We acknowledge Dan Strano for his foundational contributions to the original
Metriq platform and for initiating the early design work that later evolved into
\mgym. We thank Paul Nation and Abdullah Ash Saki for several helpful
discussions, and for providing the initial software implementation of the BSEQ
benchmark that was later adapted into \mgym. The authors thank Ryan Hill for his
work on \texttt{qBraid-SDK} and for helping with a smooth integration with
\mgym. We also thank the OriginQ team for their help in interfacing with their
devices and for supplying additional runtime for running our benchmarks. 
We thank Zichang He and Ruslan Shaydulin for their support of this project. 
The authors thank Michael Sandoval and the Quantum Computing User Program (QCUP) team at Oak Ridge National Labs for providing access to Quantinuum devices, and Enrico Rinaldi for support with 
Quantinuum job runs. 
%We also acknowledge Ruslan Shaydulin and Zichang He for their assistance in collecting EPLG data on Quantinuum.
The Metriq project benefited from continuous discussions with the
members of the Open Quantum Benchmark Committee: Alexander James Rasmusson, Amit
Jamadagni, Andrea Giachero, Ed Younis, Eduardo Henrique Matos Maschio, Frederic
Barbaresco, Justin Gage Lietz, Luke Govia, Olivia Di Matteo, Paul Nation, Peter
Groszkowski, Ryan Hill, Shannon Whitlock, and Yi-Ting (Tim) Chen.
We are also grateful to them for reviewing early drafts of this manuscript.

This work was supported
by the U.S. Department of Energy, Office of Science, Office of Advanced
Scientific Computing Research, Accelerated Research in Quantum Computing under
Award Numbers DE-SC0020266, DE-SC0025336 and DE-SC0020316 as well as by IBM
under Sponsored Research Agreement No. W1975810. The IBM benchmarks were run via
the IBM Premium account provided by Paul Nation to the Unitary Foundation team.
This research used resources of the Oak Ridge Leadership Computing Facility,
which is a DOE Office of Science User Facility supported under Contract DE-
AC05-00OR22725. NS is partly supported by the European Union via the QLASS
(Quantum Glass-based Photonic Integrated Circuits - Grant Agreement No.
101135876).

\bibliography{draft_v2} 
\bibliographystyle{naturemag}

\clearpage
\appendix

\section{Benchmark configuration reference}
\label{app:benchmark-configs}

Table~\ref{tab:benchmark-configs} consolidates the configuration parameters
for every benchmark in the \mgym suite. In code, each benchmark run is fully specified by
a JSON object containing a \texttt{benchmark\_name} field plus the parameters
listed below. The schemas are also available in the repository under
\texttt{metriq\_gym/schemas/}. To make the benchmark-weight construction in
Step (iii) in Section~\ref{sec:metriq-score} transparent, the table also records the effective benchmark scale
$\mu_b$ used in the default benchmark weights.

\begin{table*}[ht]
\centering
\scriptsize
\setlength{\tabcolsep}{3pt}
\renewcommand{\arraystretch}{1.15}

\begin{tabularx}{\linewidth}{l l l l l X}
\toprule
\textbf{Benchmark} & \textbf{Parameter} & \textbf{Value} & \textbf{Scale datapoints} & 
\textbf{Effective width} & \textbf{Description} \\
\midrule

\multirow{2}{*}{BSEQ}
& \texttt{shots} & 1000
& \multirow{2}{*}{All qubit pairs}
& \multirow{2}{*}{\makecell[l]{Suite reference\\$\mu_b = 100$}}
& Measurement repetitions per circuit \\
\cmidrule(l){2-3}\cmidrule(l){6-6}
& & & & & \textit{Auto-adapts to device connectivity; no width parameter needed.} \\
\midrule

\multirow{5}{*}{EPLG}
& \texttt{num\_samples} & 10
& \multirow{5}{*}{\makecell[l]{$n \in \{10,20,50,100\}$}}
& \multirow{5}{*}{\makecell[l]{$\mu_b \approx 72.2$}}
& Random circuit instances per chain length \\
& \texttt{num\_qubits\_in\_chain} & \textit{(varies)}
& & & Number of qubits $n$ in linear chain\\
& \texttt{shots} & 1000
& & & Measurement repetitions per circuit \\
& \texttt{lengths} & \texttt{[2,4,8,\ldots,500]}
& & & RB sequence lengths for the exponential fit \\
\cmidrule(l){2-3}\cmidrule(l){6-6}
& & & & & \textit{Reduced configs for cost-constrained platforms (see Section~\ref{sec:benchmark-eplg}).} \\
\midrule

\multirow{5}{*}{\makecell[l]{Mirror\\Circuits}}
& \texttt{width} & \textit{(varies)}
& \multirow{5}{*}{\makecell[l]{$n \in$\\$\{8, 16, 24, 32, 64, 128\}$}}
& \multirow{5}{*}{\makecell[l]{$\mu_b \approx 82.4$}}
& Number of qubits $n$ \\
& \texttt{num\_layers} & \textit{(varies)} & & & Half-circuit depth $d$ (total $\approx 2d$) \\
& \texttt{shots} & 1000 & & & Measurement repetitions per circuit \\
& \texttt{two\_qubit\_gate\_prob} & 0.5 & & & Per-layer two-qubit gate probability \\
& \texttt{num\_circuits} & 10 & & & Independent circuit samples at each $(w,d)$ \\
\midrule

\multirow{5}{*}{CLOPS}
& \texttt{num\_qubits} & 100
& \multirow{5}{*}{$n = 100$}
& \multirow{5}{*}{\makecell[l]{$\mu_b = 100$}}
& Number of qubits $n$ \\
& \texttt{num\_layers} & 100 & & & Layers per circuit \\
& \texttt{num\_circuits} & 1000 & & & Circuit instances in the workload \\
& \texttt{shots} & 100 & & & Shots per circuit \\
& \texttt{mode} & \texttt{twirled} & & & \texttt{instantiated}/\texttt{parameterized}/\texttt{twirled} \\
\midrule

\multirow{2}{*}{\makecell[l]{QML\\Kernel}}
& \texttt{num\_qubits} & \textit{(varies)}
& \multirow{2}{*}{\makecell[l]{$n\in\{10,20,30,50\}$}}
& \multirow{2}{*}{\makecell[l]{$\mu_b \approx 35.5$}}
& Number of qubits $n$ \\
& \texttt{shots} & 1000 & & & Measurement repetitions \\
\midrule

\multirow{2}{*}{WIT}
& \texttt{num\_qubits} & 7
& \multirow{2}{*}{$n = 7$}
& \multirow{2}{*}{\makecell[l]{$\mu_b = 7$}}
& Number of qubits $n$\\
& \texttt{shots} & 8192 & & & Measurement repetitions \\
\midrule

\multirow{8}{*}{LR-QAOA}
& \texttt{graph\_type} & \texttt{"1D"}
& \multirow{8}{*}{\makecell[l]{$n\in\{10,20,50,100\}$}}
& \multirow{8}{*}{\makecell[l]{$\mu_b \approx 72.2$}}
& Graph topology \\
& \texttt{num\_qubits} & \textit{(varies)} & & & Number of qubits $n$\\
& \texttt{shots} & 1000 & & & Measurement repetitions \\
& \texttt{trials} & 10 & & & Independent optimization trials \\
& \texttt{confidence\_level} & 0.999 & & & Statistical confidence threshold \\
& \texttt{num\_random\_trials} & 25 & & & Random baseline trials \\
& \texttt{qaoa\_layers} & \texttt{[10]} & & & QAOA circuit depth $p$ \\
& \texttt{delta\_beta}, \texttt{delta\_gamma} & 0.3, 0.6 & & & Linear-ramp schedule slopes (radians) \\
\midrule

\multirow{6}{*}{QFT}
& \texttt{min\_qubits} & 4
& \multirow{6}{*}{\makecell[l]{$n\in\{4,8,12,20\}$}}
& \multirow{6}{*}{\makecell[l]{$\mu_b \approx 14.2$}}
& Smallest QFT size in sweep \\
& \texttt{max\_qubits} & 20 & & & Largest QFT size in sweep \\
& \texttt{skip\_qubits} & 4 & & & Increment between sizes \\
& \texttt{max\_circuits} & 3 & & & Circuit instances per size \\
& \texttt{shots} & 1000 & & & Measurement repetitions \\
& \texttt{method} & 1 & & & Variant (1: fwd+inv, 2: encode+inv) \\

\bottomrule
\end{tabularx}

\renewcommand{\arraystretch}{1}

\caption{Configuration parameters for each benchmark in the \mgym suite.
The \texttt{benchmark\_name} field is omitted for brevity. The \emph{Value}
column lists the settings used for the results reported in this work; entries
marked \textit{(varies)} are swept over the scale datapoints shown. The
\emph{Scale datapoints} column lists the problem sizes at which each
benchmark is evaluated and scored. The \emph{Effective width}
column records the quantity $\mu_b$ used in the default benchmark weighting scheme
(Eq. \eqref{eq:effective-width}). All parameters are user-configurable.}

\label{tab:benchmark-configs}
\end{table*}

For the suite used in Table~\ref{tab:metriq-summary}, the effective benchmark
scales listed in the \emph{Effective width} column sum to
\[
\sum_{b\in B} \mu_b
= 100 + 72.2 + 82.4 + 100 + 35.5 + 72.2 + 7 + 14.2
\approx 483.5.
\]
Hence, by Eq.~\eqref{eq:benchmark-weights},
\[
w_b = \frac{\mu_b}{\sum_{c\in B}\mu_c}
= \frac{\mu_b}{483.5}.
\]
This gives the normalized example benchmark weights reported in
Table~\ref{tab:metriq-summary}, e.g.
\[
w_{\mathrm{BSEQ}} \approx \frac{100}{483.5} = 0.2069,\qquad
w_{\mathrm{EPLG}} \approx \frac{72.2}{483.5} = 0.1494,\qquad
w_{\mathrm{MC}} \approx \frac{82.4}{483.5} = 0.1703.
\]

\section{Verbatim compilation for benchmarking with symmetric circuits}
\label{app:verbatim_compilation}

This appendix summarizes the workflow for running the QML kernel benchmark on
AWS Braket using \emph{verbatim} compilation.  The core idea is to preserve the
structure of a symmetric circuit (such as the two-half inner-product circuit
defined by our benchmark) by circumventing any optimization or qubit remapping
performed by the provider's compiler. The same approach applies to other
symmetric benchmarks where barriers or circuit symmetry must be maintained.

The general procedure is as follows. First, construct the Qiskit circuit for
the QML kernel benchmark. The circuit has a barrier between two identical
halves to prevent gate cancellation across the midpoint. For a given number of
qubits~$N$, one can generate the feature map $U_{\mathrm{ZZ}}(x)$ and build the
inner-product circuit using \verb|UnitaryOverlap|, then assign a random
parameter vector. Second, choose the appropriate native gate set for the target
device. For example, Rigetti's Ankaa processors expose \verb|rx| and \verb|rz| for single-qubit rotations and \verb|iswap| as the entangling gate, while IQM's Garnet processor uses arbitrary single-qubit rotations \verb|prx| (equivalent to \verb|r| gate in Qiskit) and \verb|cz| as its two-qubit gate and IonQ devices support the native single-qubit gates \verb|gpi| and \verb|gpi2| and the two-qubit Molmer-Sorensen
gate \verb|ms|.  Given the Qiskit QML kernel benchmarking circuit
\verb|qmlk_circuit|, the Qiskit transpiler can decompose it into any chosen
basis via

\begin{tcblisting}{
  listing engine=listings,
  listing only,
  colback=gray!5,
  colframe=gray!40,
}
iqm_gateset = [`r', `cz']
transpiled_circuit = transpile(qmlk_circuit, basis_gates=iqm_gateset)
transpiled_circuit.global_phase = 0
\end{tcblisting}
or analogous basis lists for Rigetti and IonQ.
Third, convert the resulting Qiskit circuit into a Braket circuit.  This can be done using the \verb|qbraid| toolkit, which supports translation between Qiskit and Braket representations:
\begin{tcblisting}{
  listing engine=listings,
  listing only,
  colback=gray!5,
  colframe=gray!40,
}
from qbraid import transpile as qb_transpile
qmlk_circuit_braket = qb_transpile(transpiled_circuit, "braket")
\end{tcblisting}
The output \verb|qmlk_circuit_braket| contains only native gates and preserves the
circuit's structure. Fourth, wrap the translated circuit in a verbatim box.  In
the Braket SDK, a verbatim box instructs the device to execute its contents
exactly as provided, without further optimization or qubit rewiring.  The
wrapper can be created via
\begin{tcblisting}{
  listing engine=listings,
  listing only,
  colback=gray!5,
  colframe=gray!40,
}
from braket.circuits import Circuit
qmlk_circuit_verb = Circuit().add_verbatim_box(qmlk_circuit_braket)
\end{tcblisting}
When submitting the task, one should also disable automatic qubit rewiring to ensure that the qubit indices used in the circuit match the intended physical qubits:
\begin{tcblisting}{
  listing engine=listings,
  listing only,
  colback=gray!5,
  colframe=gray!40,
}
result = device.run(qmlk_circuit_verb, shots=1000, disable_qubit_rewiring=True).result()
\end{tcblisting}

Counting gates in the compiled circuit provides a consistency check.  For
example, a 20-qubit QML kernel circuit requires two entangling layers, each
entangling neighboring pairs with two controlled-NOT operations per pair.  In
total the inner-product circuit contains $2 \times 2 \times  (20-1)  = 76$
two-qubit gates.  The verbatim circuit compiled for \texttt{iqm\_emerald} likewise contained 76
\verb|cz| gates, confirming that no additional two-qubit operations were
introduced by the transpiler.  Similar gate counts can be verified for other devices.

This workflow ensures that symmetric circuits with barriers or other structural
constraints are executed faithfully on hardware that does not natively support
barriers.  By selecting the proper native gate set and employing verbatim
compilation, one isolates hardware performance from compiler behavior and
obtains reliable benchmarking results.

\section{Metriq website}
\label{app:metriq_web}
\begin{figure}
    \centering
    \includegraphics[width=0.7\linewidth]{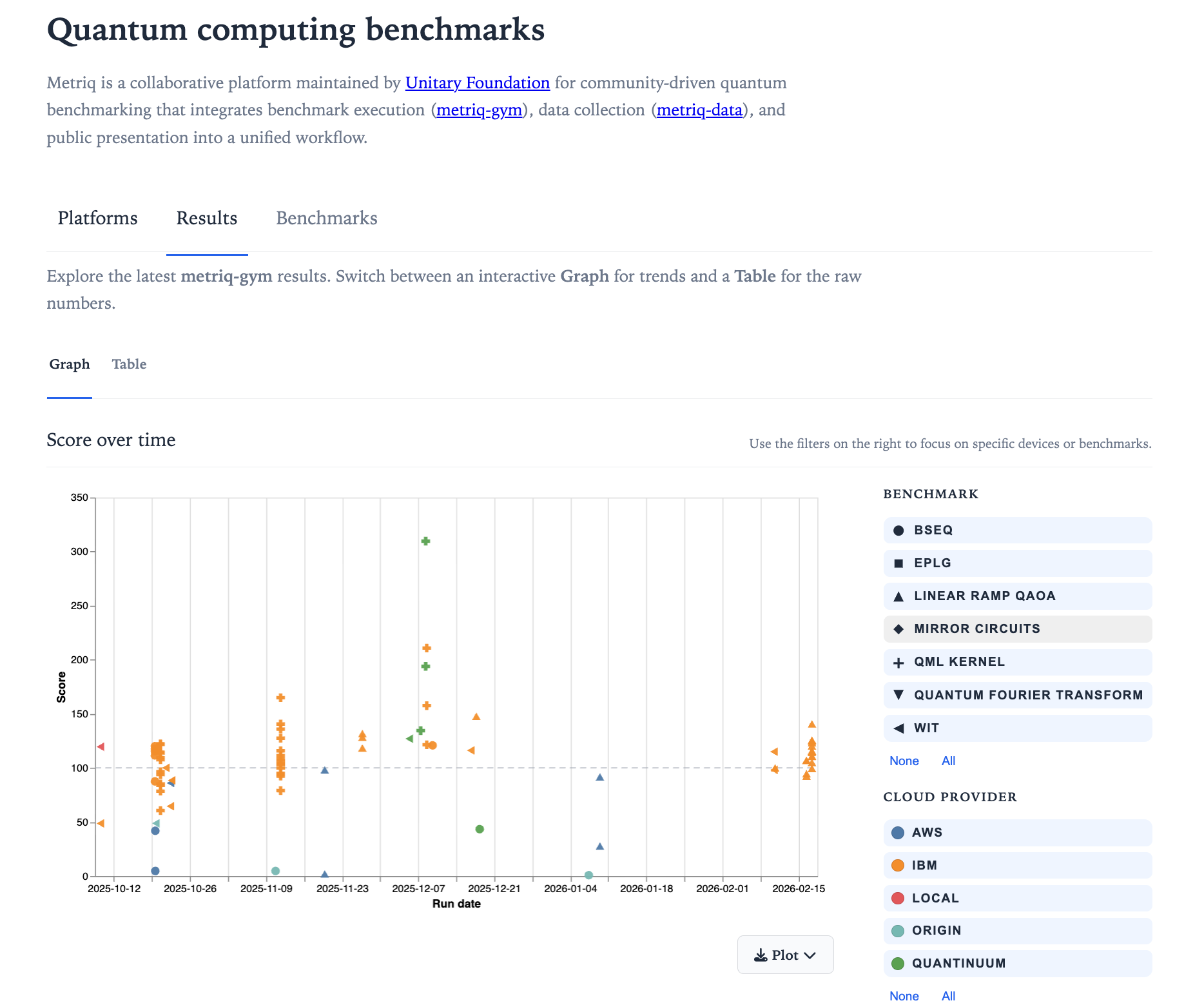}
    \caption{Front page of \hreftexttt{https://metriq.info}{metriq.info} as of February 19,
2026 (see the latest website page for updated benchmarking results), showing interactive plots for exploring benchmark results across devices and benchmarks.}
    \label{fig:screenshot_metriq_web}
\end{figure}

We provide a screenshot of the Metriq website front page as of February 19,
2026, as shown in Fig.~\ref{fig:screenshot_metriq_web}, to illustrate the user
interface of the collaborative benchmarking platform. The page presents recent
benchmark results as interactive scatter plots of score versus run date, with
options to switch between graphical and tabular views of the data. The uploaded
results also include metadata such as data taken time and parameter settings for
the benchmark. Users can filter the displayed points by benchmark type and by
provider, which enables focused comparisons of different quantum devices and
benchmark families, as well as their historical benchmark results. This figure is
intended to give a concrete view of how Metriq supports transparent,
cross-platform exploration of quantum computing benchmark results.

\end{document}